\documentclass{autart}

\usepackage[utf8]{inputenc}
\usepackage[T1]{fontenc}
\usepackage{lmodern}

\usepackage{amsmath,amssymb,amsfonts,mathtools}
\usepackage{empheq}

\usepackage{graphicx}
\usepackage{booktabs}
\usepackage{multirow}
\usepackage[table]{xcolor}

\usepackage{cite}
\usepackage{url}

\usepackage{float}
\usepackage{pifont}

\usepackage{silence}
\usepackage[compatibility=false]{caption}
\usepackage{subcaption}
\begin{document}

\begin{frontmatter}

\title{From Well-Posed Inversion to Learning Design: Physics-Informed Neural Estimation for Autonomic Regulation}

\author[addr1]{Sara Nour Sadoun}, 
\ead{maria-sara-nour.sadoun@inria.fr}
\author[addr2]{Giuseppe Alessio D'Inverno}, 
\author[addr1]{Francois Cottin}, 
\author[addr1]{Arnaud Boutin} and
\author[addr1]{Taous-Meriem Laleg-Kirati}
\ead{taous-meriem.laleg@inria.fr}

\address[addr1]{Universite Paris-Saclay, INRIA, CIAMS, 91190 Gif-sur-Yvette, France}
\address[addr2]{Laboratoire de Mathematiques d'Orsay, Universite Paris-Saclay, 91400 Orsay, France}

\begin{abstract}
Learning-based and physics-informed methods are increasingly used for inverse estimation in controlled nonlinear dynamical systems. However, in many such approaches, the theoretic requirements that make unknown-input reconstruction meaningful, namely well-posedness in the sense of Hadamard, are often disregarded or weakly addressed through generic regularization terms with no explicit guarantees. In this work, we adopt a complementary viewpoint in which these control-theoretic  and structural conditions inform the estimator design and constrain its training. We thus develop a physics-informed input-state neural estimator for joint unknown-input and state estimation in nonlinear controlled systems with partial measurements. In the present work, this general framework is instantiated on a model of autonomic cardiac regulation, provides a concrete study case. The estimator is formulated as an inverse neural map conditioned on time and measured outputs, and is trained under data fidelity and dynamical consistency constraints. To ensure it complies with the same structural requirements imposed in robust estimation, we derive left-invertibility conditions by differential-algebraic elimination and embed the resulting constraints directly into the training objective. We further analyze a priori the stability of the inverse mapping to output perturbations and derive a conservative Lipschitz bound that guides the tuning of cost functional hyper-parameters. The framework is evaluated on simulated data, where ground truth data is available, and on two distinct datasets of real cardiovascular recordings. The results show that incorporating control-theoretic solvability constraints into physics-informed learning improves the reliability of inverse inference beyond forward consistency alone.
\end{abstract}

\end{frontmatter}

\section{Introduction}
Recent advances in data-driven and learning-based methods have expanded the design space and research perspective of model-based control and estimation, enabling flexible approximations of uncertain dynamics, scalable inference from high-dimensional measurements, and improved performance in regimes where first-principles models can be restrictive, incomplete, or difficult to calibrate. In particular, neural network representations have extended classical observers and filters through learned embeddings and hybrid architectures, supporting state estimation, system identification, and control while retaining physical interpretability \cite{jin2021new, chen2023data, fiedler2025data}.\\
A prominent family within Scientific Machine Learning (SciML) is physics-informed neural networks (PINNs), which embed differential-equation residuals into training objectives to enforce physical consistency. Yet operational reliability requires consistency with control-theoretic notions such as well-posedness, stability, and identifiability, placing physics-based learning in direct contact with estimation theory and motivating the broader viewpoint of physics-informed neural estimators --- learning architectures designed explicitly for inference in dynamical systems. A central motivation for such approaches arises in inverse inference problems, where the goal is not merely to simulate trajectories forward, but to infer hidden states, parameters, or unknown inputs from partial output measurements --- for which standard PINN structures are often not adapted.\\
Control input reconstruction, also referred to as unknown-input estimation, is particularly important in biomedical engineering and related domains where inputs may be unmeasured or available only through indirect sensors, and where their recovery is essential for interpretability, retrospective analysis, and model-based monitoring. In autonomic regulation specifically, reconstructing latent regulatory signals is critical when the measurable output is a coarse proxy of underlying physiology.\\
Yet, input reconstruction is inherently constrained by the requirement of Hadamard well-posedness, as distinct inputs may generate indistinguishable outputs under the same model and measurement configuration, while small observation errors may amplify into larger differences in the reconstructed input. Without conditions ensuring that the input is uniquely and robustly determined by the observations, a learned inverse model may reproduce the observations while producing physically plausible but non-unique explanations --- a difficulty further exacerbated by the flexibility of neural networks. As a result, a physics-based learning approach may appear accurate in forward prediction while proving unreliable in inverse inference unless well-posedness is explicitly accounted for in the considered modeling and sensing setup.\\
From a methodological standpoint, the literature spans a broad spectrum, from classical observer-based reconstruction to modern learning-based inverse formulations. Unmeasured inputs have been addressed through sliding-mode designs \cite{sharma2011fault, veluvolu2009discrete}, adaptive and robust $\mathcal{H}_\infty$ observers for Lipschitz or differentiable nonlinear systems \cite{yang2013state, li2016state}, and finite-time or LMI-based estimators under constrained disturbances \cite{malikov2018synthesis, chakrabarty2017state}. In parallel, offline inverse strategies have recovered unknown nonlinear effects via subspace deconvolution, with performance contingent on orthogonality conditions \cite{bernal2023input}. Across these approaches, reconstructability and robustness typically rely on structural assumptions, such as matched uncertainties or partially specified input models, which can become restrictive in realistic applications.\\
More recent work has sought to relax these restrictions by combining model structure with data-driven flexibility. Gray-box Extended Kalman Filter schemes jointly estimate states and local dynamics \cite{kullberg2021online}, while Gaussian-process latent-force models represent unknown inputs as Gaussian processes in linear augmented state-space models \cite{nayek2019gaussian, vettori2024assessment, caglio2025joint}. PINN-based inverse-source formulations follow a similar idea, recovering unknown forcing terms under physics constraints \cite{kim2024review, barreau2021physics, chen2023physics}. Although several recent studies discuss identifiability issues in PINN-based workflows and emphasize their practical impact \cite{kharazmi2021identifiability, zhang2022analyses, haywood2025response, minadakis2024application}, explicit guarantees remain largely absent. This motivates the need for estimator architectures and training procedures that do not merely enforce physics approximately, but incorporate structural solvability constraints directly into the learning problem.\\
At the same time, framing unknown-input estimation as an inverse problem connects it to a growing literature on neural inverse solvers that address robustness and ambiguity. Deep networks have been shown to approximate inverse operators robustly under suitable restrictions on the forward operator, with broader performance characterizations developed for deep inverse models \cite{LermaPinedaPetersen2022Stable, Buskulic2024Thesis, Amjad2022Thesis}. This ambiguity has motivated invertible neural architectures and flow-based models that learn the forward map jointly with a latent-augmented inverse pass \cite{Ardizzone2018INN, Dinh2014NICE, Dinh2016RealNVP, Kingma2018Glow}. While these approaches strengthen learning-based inversion, their guarantees are typically formulated at the level of the learned mapping itself and, when model-based considerations are embedded, remain restricted to linear models; the use of physics penalties improves plausibility \cite{krishnapriyan2021characterizing} but does not guarantee unique solvability or structural consistency of the inverse problem.\\
In controlled nonlinear dynamical systems, reliability fundamentally depends on whether the underlying model and sensing configuration admit a unique and stable reconstruction.\\
Our objective is therefore to explore such a structurally constrained learning strategy for joint state and unknown-input estimation in a nonlinear controlled dynamical model of autonomic cardiac regulation proposed in \cite{ataee2010baroreflex}, in which the input enters through the model's primary stiff nonlinearity. The present case study provides a concrete and necessary setting in which the framework's assumptions, implementation, and performance can be explicitly examined, while remaining generalizable to related inverse problems.\\
With these considerations in mind, we present a new framework, a Physics-Informed Input--State Neural Estimator (PIISNE) designed on a nonlinear controlled dynamical system of autonomic cardiac regulation, with the main contributions:
\begin{itemize}
    \item[(i)] We propose an augmented estimator formulation for the inverse input reconstruction problem associated with the autonomic cardiac regulation model proposed in \cite{ataee2010baroreflex}, in which the neural representation is explicitly conditioned on the measured output and time $t$, casting the learned map as an inverse estimator of the unknown regulatory input.
    \item[(ii)] We establish left-invertibility guarantees based on differential-algebraic elimination, deriving input--output relations and constraints ensuring unique reconstructibility. Preliminary results concerning this step were developed in a first contribution \cite{sadoun2026physics}.
    \item[(iii)] We investigate the stability of the inverse I/O mapping with respect to output perturbations and derive a conservative bound of a Lipschitz-like constant that provides insight on the choice of a set of hyperparameters in the cost function.
    \item[(iv)] We constrain the training of our neural network with a loss enforcing data fidelity, model residual consistency, and left-invertibility conditions, within a subspace ensuring Output-Input stability, to mitigate spurious non-unique inversions.
    \item[(v)] We validate the resulting estimator on simulated data where ground truth is available for both input and states, evaluate on two distinct datasets of real cardiovascular recordings with substantial inter-subject diversity and intra-subject variability, and benchmark against classical estimators.
\end{itemize}

\textbf{Paper organization.} Section II introduces the preliminary theoretical notions necessary for the rest of the paper. Section III formulates the problem addressed around the case study of the autonomic cardiac regulation model. Section IV details the model-specific structural left-invertibility analysis and how it is incorporated into learning. Section V presents the stability analysis of the inverse mapping, whose numerical study is conducted in Section VII. Section VI reports the architecture framework and training process, followed by experiments and comparisons on the datasets. Discussion and conclusion are provided in Sections VIII to XI.
\section{Preliminaries: Well-posedness of inverse reconstruction}
Inverse reconstruction problems aim to recover an unknown signal (or parameter) $u$ from observed data $y$ using a forward model
\begin{equation}
\label{eq:forward_map_general}
y = \mathcal{F}(u),
\end{equation}
where $\mathcal{F}$ denotes the input-output operator. 

\textbf{\hspace{-0.5cm} Definition (Hadamard well-posedness): } \label{def:hadamard_wellposedness} Let $\mathcal{X}$ and $\mathcal{Y}$ be normed spaces, $\mathcal{X}_{\mathrm{ad}}\subseteq \mathcal{X}$ an admissible set, and $\mathcal{F}:\mathcal{X}_{\mathrm{ad}}\to\mathcal{Y}$ a forward operator.
The inverse reconstruction problem associated with $\mathcal{F}$ is said to be well-posed in the sense of Hadamard \cite{kabanikhin2008definitions} on $\mathcal{X}_{\mathrm{ad}}$ if:
\begin{enumerate}
    \item[(i)] Existence: $\forall y \in \mathcal{F}(\mathcal{X}_{\mathrm{ad}})$ there exists at least one $u\in \mathcal{X}_{\mathrm{ad}}$ s.t. $y=\mathcal{F}(u)$.
    \item[(ii)] Uniqueness: $\forall y \in \mathcal{F}(\mathcal{X}_{\mathrm{ad}})$ there exists at most one $u\in \mathcal{X}_{\mathrm{ad}}$ such that $y=\mathcal{F}(u)$ ($i.e.$, $\mathcal{F}$ is injective on $\mathcal{X}_{\mathrm{ad}}$).
    \item[(iii)] Stability: The inverse mapping $\mathcal{R}=\mathcal{F}^{-1}:\mathcal{F}(\mathcal{X}_{\mathrm{ad}})\to\mathcal{X}_{\mathrm{ad}}$ is continuous with respect to the chosen norms.
\end{enumerate}

In practical terms, the stability requirement in Definition~\ref{def:hadamard_wellposedness} is often strengthened to a Lipschitz-type dependence on the data, which provides an explicit conditioning constant.

\textbf{\hspace{-0.5cm}\textbf{Definition (left-invertibility and input/output mapping)} \cite{hirschorn1979invertibility}: } Consider the dynamical system
\begin{equation}
    \dot{x}(t)=f\!\big(x(t),u(t)\big), 
    \qquad 
    y(t)=h \!\big(x(t),u(t)\big),
\end{equation}
where $x(t)\in\mathbb{R}^n$ is the state, $u(t)\in\mathbb{R}$ is an input (hereafter regarded as unknown), and $y(t)\in\mathbb{R}$ is the measured output. Assume $f,h\in\mathcal{C}^k$ for some $k\in\mathbb{N}$.

Fix $x(0)=x_0$. For any admissible input $u(\cdot)$, denote the corresponding output trajectory by $y(\cdot;x_0,u)$ and define the input--output map
\[
\mathcal{F}_{x_0}:\ u(\cdot)\ \longmapsto\ y(\cdot;x_0,u).
\]
The system is \emph{left-invertible} (around $x_0$) if there exists a finite $T>0$ such that
\[
\mathcal{F}_{x_0}(u_1)=\mathcal{F}_{x_0}(u_2)
\quad\Longrightarrow\quad
u_1(t)=u_2(t),\ \forall t\in[0,T].
\]
Equivalently, for the same initial condition, two distinct inputs cannot generate the same output trajectory over $[0,T]$.

The left-invertibility of finite-dimensional systems described by ODE models has been extensively discussed \cite{tsiantis2018optimality, hirschorn1979invertibility, chakrabarty2017state, engelhardt2016learning, fonod2014a}. However, most contributions tend to linear systems or input-affine nonlinear systems, whereas in other settings—such as the present model—the input enters the dynamics only through nonlinear terms.\\
A closely related line of work concerns structural identifiability \cite{denis2004equivalence, wanika2024structural}. Algebro-differential elimination methods \cite{chis2011structural}, including the Rosenfeld--Gr\"obner algorithm \cite{hubert2000factorization}, make it possible to eliminate the latent state $x$ and derive an input-output relation expressed solely in terms of $u$, $y$, and their time derivatives. The existence and uniqueness properties of solutions to such relations can then be used to examine injectivity of the input-output mapping, provide us with a practical tool to assess left-invertibility.\\

\textbf{\hspace{-0.5cm} Definition (Lipschitz stability and conditioning constant): } \label{def:lipschitz_stability} The inverse reconstruction is said to be Lipschitz stable on $\mathcal{X}_{\mathrm{ad}}$ if there exists $L>0$ s.t. $\forall u_1,u_2\in\mathcal{X}_{\mathrm{ad}}$,
\begin{equation}
\label{eq:hadamard_stability_general}
\|u_1-u_2\|_{\mathcal{X}} \;\le\; L \,\|\mathcal{F}(u_1)-\mathcal{F}(u_2)\|_{\mathcal{Y}}.
\end{equation}

For dynamical inverse problems, reconstruction is performed over a finite time window $W=[0,T]$ (or a discrete grid $W=\{t_k\}_{k=1}^N$), and stability may depend on norms that include derivatives of the measured output. Accordingly, a stability requirement may take the form
\begin{equation}
\label{eq:windowed_stability_general}
\|u_1-u_2\|_{\infty,W} \;\le\; L \, \|y_1-y_2\|_{\mathcal{Y}_W},
\end{equation}
where $\mathcal{Y}_W$ can be chosen to encode the regularity required by the reconstruction procedure (e.g., $\mathcal{Y}_W = W^{k,\infty}(W)$ for some $k\ge 0$).

\noindent Stability typically fails in inverse problems, motivating regularization theory on stability modulus that is strongly model—and measurement-dependent \cite{engl2007inverse}. Recent work explicit stability requirements when inverse maps are approximated with learning-based solvers. \cite{LermaPinedaPetersen2022Stable} analyzes the Lipschitz stability of a restricted finite-dimensional inverse problem and analyzes the noise robustness of neural reconstructions under this Lipschitz inverse setting. In PINN setting, \cite{de2024numerical} assumes conditional stability of the inverse PDE problem.
% \noindent Further in this work, we address stability of inverse I/O reconstruction in a non-affine controlled, nonlinear ODE setting by deriving an explicit Lipschitz-type stability bound that guides the selection of inequalities margins and regularization strengths, and interpret the numerical conditioning behavior of the inverse input reconstruction.

\section{Problem Formulation}
\subsection{Physiology-based mathematical model of autonomic cardiac regulation}

We model the interaction between mean arterial blood pressure (BP) and heart rate (HR) through a low-dimensional baroreflex-driven dynamical system. The formulation used throughout this paper follows \cite{ottesen1997modelling}, which proposes a coupled nonlinear ODE model. The present work considers the model without explicit delays, as \cite{ottesen1997modelling} discusses delayed mechanisms in connection with the appearance of physiological oscillations as Mayer waves.

\begin{figure}[htbp]
    \centering
    \includegraphics[width=0.78\linewidth]{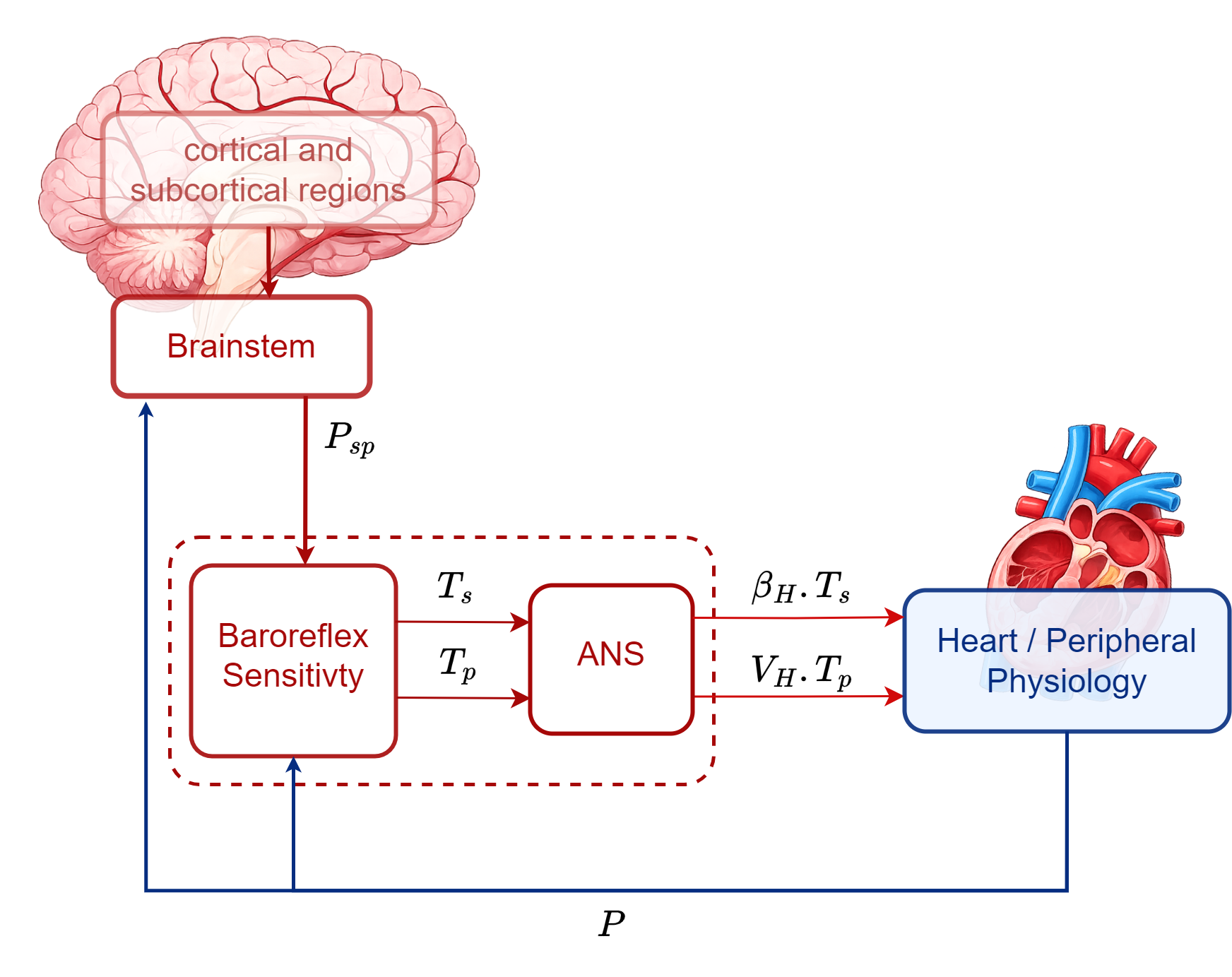}
    \caption{Neural control of cardiovascular function via the baroreflex arc}
    \label{fig:BAROREF}
\end{figure}

Let $H(t)$ denote HR and $P(t)$ denote mean arterial BP. We introduce the state vector $x(t)=[H(t),\,P(t)]^\top$ and take the pressure set-point $P_{sp}(t)$ as input, i.e., $u(t)=P_{sp}(t)$. The dynamics can be written as follows
\begin{equation}
\left\{
    \begin{aligned}
        \dot{H}(t)=f_H(x(t), u(t))\\
        \dot{P}(t)=f_P(x(t), u(t)),
    \end{aligned} \right. 
\end{equation}
where $x$ denotes the state vector $x=[H,\,P]$ and $u$ the system's input $u=P_{sp}$, and:
{\footnotesize 
\begin{subequations}\label{eq:hpta}
\vspace{-0.6cm}
\begin{empheq}[left=\empheqlbrace]{align}
f_H(x, u)
&=\frac{\beta_H T_s
}{1+\gamma T_p}-V_H T_p+\delta_H\left(H_0-H\right),
\label{eq:hta}\\
f_P(x, u)
&=-\frac{P}{R_a\left(1+\alpha T_s\right) C_a}+\frac{H \Delta V}{C_a}.
\label{eq:pta}
\end{empheq}
\end{subequations}}
The coupling between $P(t)$ and $H(t)$ is closed through two complementary autonomic drives. Baroreflex activation is represented by the sigmoid
\[
\sigma(P(t))=\frac{1}{1+e^{-\alpha_{sp}\left(P(t)-P_{sp}(t)\right)}},
\]
from which the parasympathetic activity is defined as $T_p=\sigma(P(t))$ and the sympathetic activity as $T_s=1-\sigma(P(t))$. In this representation, $P_{sp}(t)$ acts as a reference around which the baroreflex term modulates the autonomic balance.\\
\noindent The use of sigmoid $\sigma(P)$ functionals is a standard choice within the physiological modeling community to represent a smooth, saturating nonlinearity: the response increases progressively with $P(t)$, transitions around a characteristic threshold, and approaches an asymptote.\\
All parameters in \eqref{eq:hpta} are physiologically interpretable and are reported in Table~\ref{tab:parameters}.
\begin{table}[htbp]
\centering
\resizebox{0.93\linewidth}{!}{%
\begin{tabular}{l l l}
\hline
\textbf{Parameter} & \textbf{Definition} & \textbf{Nominal Value} \\ \hline
$R_a$  & minimum arterial resistance              & 0.6 mmHg$\cdot$g$\cdot$ml$^{-1}$ \\
$C_a$    & arterial compliance                      & 1.55 mmHg$\cdot$g$^{-1}$ \\
$\Delta V$ & stroke volume                           & 50 ml \\
$\beta_H$ & sympathetic control of HR                & 0.84 s$^{-2}$ \\
$\gamma$  & parasympathetic damping of $\beta_H$     & 0.2 \\
$V_H$     & parasympathetic control of HR            & 1.17 s$^{-1}$ \\
$\delta_H$& relaxation time                          & 1.7 s$^{-1}$ \\
$H_0$     & intrinsic HR                             & 100 min$^{-1}$ \\
$\alpha$  & sympathetic effect on $R_a$              & 1.3 \\
$\alpha_{sp}$ & Slope of sigmoid determining baroreflex response & 0.07 mmHg$^{-1}$\\
\hline
\end{tabular}%
}
\vspace{0.3cm}
\caption{Parameter definitions and nominal values.}
\vspace{-0.2cm}
\label{tab:parameters}
\end{table}
A schematic representation of the baroreflex arc and the associated interactions captured by the model is shown in Figure~\ref{fig:BAROREF}.
\subsection{Problem formulation}
The problem considered in this work is broadly cast as a joint unknown input-state estimation problem for nonlinear controlled dynamical systems, in which an unknown driving signal must be jointly reconstructed with unmeasured state components from partial and noisy output observations. Accordingly, we consider a controlled nonlinear dynamical system of the form
\begin{equation}
    \dot{x}(t)=f(x(t),u(t)), \qquad y(t)=C\,x(t),
\end{equation}
where $x(t)\in\mathbb{R}^n$ denotes the state, $u(t)$ an unknown input to be reconstructed, and $y(t)$ the measured output. $\hat x(t), \hat y(t), \hat u(t)$ denote their estimation. Given noisy measurements $y^{\mathrm{m}}(t)$ on a finite time horizon $[0,T]$, the associated joint state--input recovery problem can be formulated as
\begin{equation}\label{eq:joint_reconstruction_stable}
\begin{aligned}
\min_{x(\cdot),\,u(\cdot)} \quad
& \underbrace{\int_{0}^{T}\!\!\|y^{\mathrm{m}}(t)-C\hat{x}(t)\|^{2}\,dt}_{J_{\mathrm{data}}} \\
&\quad + \lambda_{\mathrm{dyn}}
\underbrace{\int_{0}^{T}\!\!\|\dot{x}(t)-f(\hat{x}(t),\hat{u}(t))\|^{2}\,dt}_{J_{\mathrm{dyn}}} \\
&\quad + \lambda_{\mathrm{LI}}\,J_{\mathrm{LI}}\big(\hat{y}(t),\hat{u}(t)\big) \\
\text{s.t.}\quad
& (x(\cdot),u(\cdot)) \in \mathcal{A}_{\mathrm{stab}}.
\end{aligned}
\end{equation}
where $J_{data}$  accounts for data fidelity, and $J_{dyn}$ for physical consistency. $J_{\mathrm{LI}}$ denotes a term constraining the optimization landscape to solutions consistent with left-invertibility conditions, thereby promoting uniqueness of the recovered input, and $\mathcal{A}_{\mathrm{stab}}$ denotes an admissible set enforcing stability of the reconstruction with respect to perturbations in the measured output over $[0,T]$.

This formulation naturally motivates a physics-informed input--state neural estimator, obtained by parameterizing $x(\cdot)$ and $u(\cdot)$ and minimizing \eqref{eq:joint_reconstruction_stable} under dynamical consistency.

\hspace{-0.5cm} \textbf{Study case (Autonomic cardiac regulation):} The present framework is investigated for a physiological study case of practical interest: the estimation of central (neural) regulation acting on the cardiovascular system through modulation of the baroreflex reference. In the considered model, this effect is represented by the pressure set-point $P_{\mathrm{sp}}(t)$, defined as an interoceptive physiological command shaping the hemodynamic response $(H(t),P(t))$.

Accordingly, let the unknown input be $u(t)=P_{\mathrm{sp}}(t)$; the state, $x(t)=[H(t),\,P(t)]^\top$, and the measured output be the heart rate:
\begin{equation}
    y(t)=C\,x(t)=H(t),
    \qquad
    C=\begin{bmatrix}1 & 0\end{bmatrix}.
\end{equation}
The objective is therefore to reconstruct the unknown neural command $P_{\mathrm{sp}}(t)$ while simultaneously estimating the unmeasured arterial pressure state $P(t)$ from noisy heart-rate measurements under the coupled nonlinear physiology-based dynamics in \eqref{eq:hpta}.
\section{Left-invertibility of the autonomic cardiac regulation model}
\noindent This section derives an explicit \emph{triangular} differential--algebraic input--output representation for the model \eqref{eq:hpta}. The resulting characterization provides structural conditions ensuring that the unknown control input $P_{\mathrm{sp}}(t)$ is structurally identifiable and that the system is left-invertible with respect to $P_{\mathrm{sp}}(\cdot)$.\\
\noindent The goal is to obtain conditions that can be used to constrain training so that the reconstruction problem is well-posed. Unlike approaches that augment the state by postulating a parametric dynamics for the unknown input, we seek to reconstruct $P_{\mathrm{sp}}(t)$ without imposing any evolution law on it, consistent with its interpretation as an interoceptive physiological command.

\noindent \textbf{\textit{Remark.}} For readability, time dependence is omitted throughout this section; overdots denote differentiation with respect to time (e.g., $\dot{x}=\frac{d}{dt}x(t)$).

\noindent \textbf{Proposition.} The model \eqref{eq:hta}--\eqref{eq:pta} admits the following triangular differential--algebraic input--output mapping:
\begin{equation}\label{eq:system_Psp_Q_sdot}
\begin{cases}
\dot P_{\mathrm{sp}} + a(s)\, P_{\mathrm{sp}}
= -\,\dfrac{1}{\alpha_{sp}}\Bigg(\dfrac{\dot s}{s(1-s)}
+ a(s)\,\log\!\Big(\dfrac{s}{1-s}\Big)\Bigg) + b\,H,\\[0.25cm]
Q(s,Y) = V_H \gamma s^2 + s\big(\beta_H + V_H + \gamma Y\big) + (Y - \beta_H) = 0,\\[0.25cm]
\dot s = -\,\dfrac{\gamma s + 1}{D(s,Y)}\,\dot Y,
\end{cases}
\end{equation}
\begin{equation}\label{eq:a_b_defs}
\text{where } 
s := T_p,\quad
a(s) := \frac{1}{R_a C_a\,[\,1+\alpha(1-s)\,]}, \quad b := \frac{\Delta V}{C_a}, \end{equation}
\begin{equation}
D(s,Y)=2V_H \gamma\,s + \beta_H + V_H + \gamma Y \quad
\text{, } Y := \dot H - \delta_H\,(H_0 - H).
\end{equation}
An admissible control input $P_{\mathrm{sp}}$ that generates the output $H$ uniquely exists, provided that
\begin{equation}\label{eq:identif} (C_{LI});
\begin{cases}
D(s,Y) \neq 0,\\[2pt]
-V_H \le Y \le \beta_H,\\[2pt]
s\notin\{0,1\}.
\end{cases}
\end{equation}
Once $s$ is determined from $H$, the coupling between $P$ and $P_{\mathrm{sp}}$ allows transferring a boundary/initial condition between $P_{\mathrm{sp}}$ and $P$, which is useful to avoid any dynamical hypothesis on the unmeasurable $P_{\mathrm{sp}}$.

\noindent \textbf{\textit{Proof.}}
\noindent Starting from \eqref{eq:pta}, set $ 
s := T_p \in [0,1], \,\, T_s = 1-T_p = 1-s.$

Substituting into \eqref{eq:pta} and using the definitions in \eqref{eq:a_b_defs} yields the compact form
\begin{equation}\label{eq:Pdot_compact}
\dot P = -\,a(s)\,P + b\,H.
\end{equation}
Next, invert the logistic relation that defines $s$:
\[
s = \frac{1}{1 + e^{-\alpha_{sp}(P - P_{\mathrm{sp}})}}
\quad\Longrightarrow\quad
P = P_{\mathrm{sp}} + \frac{1}{\alpha_{sp}}\log\!\Big(\frac{s}{1-s}\Big),
\]
\[\quad s\notin\{0,1\}.\]
Substituting this expression for $P$ into \eqref{eq:Pdot_compact} gives
\begin{equation}\label{eq:Psp_dae}
\dot P_{\mathrm{sp}} + a(s)\, P_{\mathrm{sp}}
= -\,\frac{1}{\alpha_{sp}}\Big(\frac{\dot s}{s(1-s)}
+ a(s)\,\log\!\Big(\frac{s}{1-s}\Big)\Big)
+ b\,H,
\end{equation}
which provides a differential relation for $P_{\mathrm{sp}}$ once $s$ and $\dot s$ are determined from $H$.

\smallskip
\noindent To determine $s$ algebraically, rewrite \eqref{eq:hta} with $T_p=s$ and $T_s=1-s$, and introduce
\[
Y := \dot H - \delta_H\,(H_0 - H).
\]
Then \eqref{eq:hta} is equivalent to the quadratic constraint
\begin{equation}\label{eq:Q}
Q(s,Y) = V_H \gamma s^2 + s\big(\beta_H + V_H + \gamma Y\big) + (Y - \beta_H)=0.
\end{equation}
Differentiating \eqref{eq:Q} along trajectories yields
\[
\frac{dQ(s,Y)}{dt}=\frac{\partial Q}{\partial s}\,\dot s+\frac{\partial Q}{\partial Y}\,\dot Y = 0,
\]
with $\frac{\partial Q}{\partial Y}=\gamma s+1$, and
\[
D(s,Y)=\frac{\partial Q}{\partial s}=2V_H\gamma s+\beta_H+V_H+\gamma Y.
\]
\begin{equation}\label{eq:sdot_explicit}
\text{Hence:}\qquad\qquad \dot s
= -\,\frac{\gamma s + 1}{D(s,Y)}\,(\ddot H + \delta_H \dot H),
\end{equation}

which is well-defined under $D(s,Y) \neq 0$. Collecting \eqref{eq:Psp_dae}, \eqref{eq:Q}, and \eqref{eq:sdot_explicit} yields \eqref{eq:system_Psp_Q_sdot}. 

\subsection*{Admissibility of the algebraic solution $s$}
\noindent We now study the existence of an admissible solution $s\in(0,1)$ to \eqref{eq:Q}. Recall \eqref{eq:Q}:
\begin{equation}
Q(s,Y)= V_H \gamma \, s^2 + s\big(\beta_H+V_H+\gamma Y\big) + (Y-\beta_H)=0.
\end{equation}
Set $ \quad A:=V_H \gamma,\quad B:=\beta_H+V_H+\gamma Y,\quad C:=Y-\beta_H,$
 so that $Q(s,Y)=As^2+Bs+C$.

The discriminant is
\begin{equation}\label{eq:disc}
\Delta = B^2 - 4AC
= (\beta_H+V_H+\gamma Y)^2 - 4(V_H \gamma)(Y-\beta_H).
\end{equation}

\noindent \textbf{$1-$ Sign structure.}
Let $r_1,r_2$ be the roots of $Q(s,Y)=0$. By Vi\`ete's formulas,
\[
r_1+r_2 = -\frac{B}{A},\qquad r_1r_2 = \frac{C}{A}.
\]
Since $A>0$:
\begin{itemize}
\item If $C\le 0$ (equivalently $Y\le \beta_H$), then $r_1r_2\le 0$ and the roots have opposite signs; thus, there exists exactly one nonnegative root whenever $\Delta\ge 0$.
\item If $C>0$ (equivalently $Y>\beta_H$), then $B=\beta_H+V_H+\gamma Y>0$, implying $r_1+r_2<0$ and $r_1r_2>0$; therefore, both roots are negative.
\end{itemize}

\noindent \textbf{$2-$ Root in the interval $(0,1)$.}
A sufficient condition for the existence of a root in $(0,1)$ is
\begin{equation}\label{eq:IVT_check}
Q(0)\,Q(1)\le 0.
\end{equation}
Here,$\qquad
Q(0)=Y-\beta_H,\qquad
Q(1)=V_H\gamma + (\beta_H+V_H+\gamma Y) + (Y-\beta_H)=(1+\gamma)(V_H+Y).$

Therefore, \eqref{eq:IVT_check} is equivalent to $
\qquad (Y-\beta_H)\,(V_H+Y)\le 0
\qquad\Longleftrightarrow\qquad
-V_H \le Y \le \beta_H,$
which yields the admissibility condition stated in \eqref{eq:identif}.

\section{Stability and conditioning of the inverse I/O reconstruction}

\noindent A central viewpoint in this work is that learning the inverse map is viewed as an estimator design problem rather than a generic function fitting. The designed estimator aims to drive \(H_{\theta}(t)\to H_{\mathrm{data}}(t)\). Yet, even when \(H_{\theta}\) closely matches \(H_{\mathrm{data}}\), uniqueness alone does not guarantee accurate reconstruction: the inverse map may be ill-conditioned, so that small perturbations in \(H\) (and its derivative estimates) are amplified into large errors on \(u\).\\
This motivates an explicit stability analysis of the inverse reconstruction on \([0,T]\), in which we derive a Lipschitz-type bound with a constant depending on the model parameters and the admissibility margins. In particular, we make precise how the admissible region enforced by our constraints—specifically, sigmoid non-saturation and a non-vanishing denominator—controls conditioning and guides the choice of constraint margins and regularization strengths for robust input estimation.

\subsection{Objective definition (Output-Input Stability)}

Fix a time window $I=[0,T]$ (or $I=\{t_1,\dots,t_N\}$). We study the stability of the
\emph{inverse operator} that maps a heart-rate output trajectory $y(\cdot)=H(\cdot)$ to an input trajectory
$u(\cdot)=P_{sp}(\cdot)$.

\noindent We consider two admissible trajectories $(H_1,u_1)$ and $(H_2,u_2)$ and aim to prove a Lipschitz-type stability bound:
\[
\|u_1-u_2\|_{\infty,I}\ \le\ C_H(\delta)\,\|y_1-y_2\|_{W^{2,\infty}(I)},
\]
where \(C_H(\delta)\) depends on margin parameters $\delta$ and on admissibility margins \(\delta\),
and: \\$
\|y_1-y_2\|_{W^{2,\infty}(I)}
= \|y_1-y_2\|_{\infty,I}
+\|\dot y_1-\dot y_2\|_{\infty,I}
+\|\ddot y_1-\ddot y_2\|_{\infty,I}.
$ and $\|f\|_{\infty,W}:=\sup_{t\in I}|f(t)|.$

\subsection{Stability constant}
% ---------- Notation blocks (place before the theorem) ----------
% \newcommand{\Params}{\eta}
\newcommand{\Margins}{\delta}

\paragraph*{Theorem (Hadamard-type stability of the inverse I/O map).}
\label{thm:inverse_io_stability}

Let \(T>0\) and let \(I=[0,T]\). Let \(H_1,H_2\in W^{2,\infty}(I)\) and define, for \(i\in\{1,2\}\),
\begin{equation}
\label{eq:Y_def_thm_clean}
Y_i(t):=\dot H_i(t)+\delta_H\big(H_i(t)-H_0\big),\qquad t\in I.
\end{equation}
% The inverse I/O mapping in \eqref{eq:Psp_dae} is written as:
% \begin{equation}
% \label{eq:system_Psp_Q_sdot}
% \dot u_i+a(s_i)u_i=F(H_i,s_i,\dot s_i),
% \end{equation}
% \begin{equation}
% \label{eq:F_def_correct}
% \text{with }F(H,s,\dot s):= bH -\frac{1}{\alpha_{sp}} \left(\frac{\dot s}{s(1-s)}
% + a(s)\log\frac{s}{1-s}\right),
% \end{equation}

\textbf{Admissibility.}
We denote \(\delta=\{\varepsilon_{\mathrm{sat}},\varepsilon_{\mathrm{deg}}\}>0\) and assume that
for all \(t\in I\) and \(i\in\{1,2\}\),
\begin{equation}
\label{eq:admissibility_thm_clean}
s_i(t)\in[\varepsilon_{\mathrm{sat}},1-\varepsilon_{\mathrm{sat}}],
\qquad |D(s_i(t),Y_i(t))|\ge \varepsilon_{\mathrm{deg}},
\end{equation}
where \(D\) is the non-degeneracy term arising from the left-invertibility relation.

\textbf{Boundedness.}
Assume \(\beta=\{U,M_{\dot Y},M_{\dot s}\}\) is finite on \(I\), with
\[
U\ge \|u_2\|_{\infty,I}, \qquad M_{\dot Y}\ge \|\dot Y_2\|_{\infty,I},
\qquad M_{\dot s}\ge \|\dot s_2\|_{\infty,I}.
\]
Then there exists an explicit constant \(C_H(\delta,\beta,T)>0\), depending only on the admissibility
margins, the boundedness quantities, the model parameters, and the window length \(T\), such that
\begin{equation}
\label{eq:main_stability_thm_clean}
\|u_1-u_2\|_{\infty,I}
\le C_H(\delta,\beta,T) \|H_1-H_2\|_{W^{2,\infty}(I)}.
\end{equation}
An admissible conservative expression for \(C_H\) is of the form:
\[
C_H:=T\Bigg[L_aU\,K_{s,H}+|b|+\frac{1}{|\alpha_{sp}|}\Big(L_{\log}K_{\dot s}+K_sK_{s,H}\Big)\Bigg],
\]
where the variables are explicit functions of $(\delta, \beta_W, T)$ given in the proof.

\textbf{Proof.}
We outline the derivation of \eqref{eq:main_stability_thm_clean} and make explicit how each
constant enters the final bound.

\medskip
\noindent\textbf{Step 0 (\(H\rightarrow Y\) is Lipschitz).}

Let $\Delta f:=f_1-f_2 \quad $ for any functional or variable $f$.

With \eqref{eq:Y_def_thm_clean}, we have:
\begin{equation}
\label{eq:deltaY_bound}
\text{Therefore, }\qquad\|\Delta Y\|_{\infty,I} \le\|\Delta\dot H\|_{\infty,I}+\delta_H\|\Delta H\|_{\infty,I}.
\end{equation}
Since \(\delta_H>1\), we obtain the conservative bound
\begin{equation}
\label{eq:deltaY_bound_delta}
\|\Delta Y\|_{\infty,I}\le\delta_H\|\Delta H\|_{W^{2,\infty}(I)}.
\end{equation}
And similarly:
\begin{equation}
\label{eq:deltaYdot_bound}
\|\Delta\dot Y\|_{\infty,I}
\le\|\Delta\ddot H\|_{\infty,I}
+\delta_H\|\Delta\dot H\|_{\infty,I}
\le\delta_H\|\Delta H\|_{W^{2,\infty}(I)}.
\end{equation}

\noindent\textbf{Step 1 (Bounding \(\Delta s\) and \(\Delta\dot s\)).}

Let the algebraic left-invertibility relation be written as
\[
Q(s,Y)=0.
\]
Denote $D:=\frac{\partial Q}{\partial s},
\qquad Q_Y:=\frac{\partial Q}{\partial Y}.$
By the admissibility condition \eqref{eq:admissibility_thm_clean}, \(|D|\ge\varepsilon_{\mathrm{deg}}\)
on the admissible set.

The implicit function theorem yields the pointwise sensitivity
\begin{equation}
\label{eq:dsdY}
\frac{\partial s}{\partial Y}(s,Y) =-\frac{Q_Y(s,Y)}{D(s,Y)}.
\end{equation}
\begin{equation}
\label{eq:Gamma_def_proof}
\text{Let}\quad \Gamma(\delta)
:=\sup_{(s,Y)\in\mathcal A_{\mathrm{stab}}}|Q_Y(s,Y)|,
\quad
L_{sY}(\delta):=\frac{\Gamma(\delta)}{\varepsilon_{\mathrm{deg}}},
\end{equation}
where
\[
\mathcal A_{\mathrm{stab}}:=\left\{(s,Y):s\in[\varepsilon_{\mathrm{sat}},1-\varepsilon_{\mathrm{sat}}], \ |D(s,Y)|\ge\varepsilon_{\mathrm{deg}}
\right\}.
\]
Then \eqref{eq:dsdY} and \eqref{eq:deltaY_bound_delta} imply
\begin{equation}
\label{eq:delta_s_bound}
\|\Delta s\|_{\infty,I}
\le
L_{sY}(\delta)\|\Delta Y\|_{\infty,I}
\le
L_{sY}(\delta)\delta_H
\|\Delta H\|_{W^{2,\infty}(I)}.
\end{equation}
\begin{equation}
\label{eq:KsH_explicit}
\text{Hence, one may take}\qquad K_{s,H}(\delta):=L_{sY}(\delta)\delta_H,
\end{equation}
\[\text{so that: }\qquad \|\Delta s\|_{\infty,I}\le K_{s,H}(\delta)\|\Delta H\|_{W^{2,\infty}(I)}.
\]

To control \(\Delta\dot s\), differentiate \(Q(s(t),Y(t))=0\) in time:
\[
D(s,Y)\dot s(t)+Q_Y(s,Y)\dot Y(t)=0.
\]
\begin{equation}
\label{eq:implicit_diff_time}
\text{\hspace{-0.3cm}Thus, } \dot s(t)=-A(s,Y)\dot Y(t),\quad A(s,Y):=\frac{Q_Y(s,Y)}{D(s,Y)}.
\end{equation}
On the admissible set, $|A(s,Y)|\le L_{sY}.$
Moreover, assume that \(A\) is Lipschitz on the admissible set:
\begin{equation}
\label{eq:boundAsubA}
|A(s_1,Y_1)-A(s_2,Y_2)| \le L_A\Big(|s_1-s_2|+|Y_1-Y_2|\Big),
\end{equation}
\begin{equation}
\label{eq:LA}
\text{where an admissible choice is }
L_A:=\sup_{\mathcal A_{\mathrm{stab}}} \left(|\partial_sA|+|\partial_YA|\right).
\end{equation}
We now write
\[
\dot s_1-\dot s_2=-A_1\dot Y_1+A_2\dot Y_2=-A_1(\dot Y_1-\dot Y_2)-(A_1-A_2)\dot Y_2.
\]
Taking \(\|\cdot\|_{\infty,I}\) and using \eqref{eq:Gamma_def_proof},
\eqref{eq:delta_s_bound}, and \eqref{eq:boundAsubA}, yields
\begin{equation}
\label{eq:sdot_diff}
\|\Delta\dot s\|_{\infty,I} \le L_{sY}\|\Delta\dot Y\|_{\infty,I}+L_AM_{\dot Y}\Big(\|\Delta s\|_{\infty,I} +\|\Delta Y\|_{\infty,I}\Big).
\end{equation}
Together with \eqref{eq:deltaY_bound_delta}, \eqref{eq:deltaYdot_bound}, and
\eqref{eq:delta_s_bound}, we obtain
\begin{equation}
\label{eq:delta_sdot_bound}
\|\Delta\dot s\|_{\infty,I}
\le K_{\dot s}(\delta,\beta)
\|\Delta H\|_{W^{2,\infty}(I)},
\end{equation}
where an admissible conservative choice is
\begin{equation}
\label{eq:Ksdot_explicit}
K_{\dot s}:=L_{sY}\delta_H+L_AM_{\dot Y}\Big(K_{s,H}+\delta_H\Big).
\end{equation}
\noindent\textbf{Step 2 (Bounding the forcing difference).}

We use the forcing term from \eqref{eq:Psp_dae}:
\[
F(H,s,\dot s):=bH-\frac{1}{\alpha_{sp}}
\left(\frac{\dot s}{s(1-s)}+a(s)\log\frac{s}{1-s}\right).
\]
Since the following estimates are taken in absolute value, the sign of the second term does
not affect the bound.
\[\text{Define }
L_{\log}(\delta) :=\frac{1}{\varepsilon_{\mathrm{sat}}(1-\varepsilon_{\mathrm{sat}})},
\quad
M_{\log}(\delta):=\log\frac{1-\varepsilon_{\mathrm{sat}}}{\varepsilon_{\mathrm{sat}}},
\]
\[
\text{and }L_{\mathrm{frac2}}(\delta):=\max_{s\in[\varepsilon_{\mathrm{sat}},1-\varepsilon_{\mathrm{sat}}]}\left|\frac{d}{ds}\frac{1}{s(1-s)}\right| \le \frac{1}{\varepsilon_{\mathrm{sat}}^2(1-\varepsilon_{\mathrm{sat}})^2}.
\]

\vspace{-0.5cm}
\[\text{Also set }\,\,
a_{\max}:=\max_{s\in[\varepsilon_{\mathrm{sat}},1-\varepsilon_{\mathrm{sat}}]}a(s),
\,\,
L_a:=\max_{s\in[\varepsilon_{\mathrm{sat}},1-\varepsilon_{\mathrm{sat}}]}|a'(s)|.
\]
For the term \(\dot s/[s(1-s)]\), we have
\[
\left|\frac{\dot s_1}{s_1(1-s_1)}-\frac{\dot s_2}{s_2(1-s_2)}\right|\le L_{\log}|\dot s_1-\dot s_2|+M_{\dot s}L_{\mathrm{frac2}}|s_1-s_2|.
\]
For the term \(a(s)\log\frac{s}{1-s}\), we have
\vspace{-0.3cm}
\begin{align*}
\left|
a(s_1)\log\frac{s_1}{1-s_1} -a(s_2)\log\frac{s_2}{1-s_2} \right|\le \\  \le \left(L_aM_{\log}+a_{\max}L_{\log}\right)|s_1-s_2|.
\end{align*}
\vspace{-1.6cm}

\begin{equation}
\label{eq:Ks}
\text{Define }K_s:=M_{\dot s}L_{\mathrm{frac2}}+L_aM_{\log}+a_{\max}L_{\log}.
\end{equation}
Combining the latter inequalities yields
\begin{align}
\label{eq:F_diff}
\|F_1-F_2\|_{\infty,I}\le{}& |b|\|H_1-H_2\|_{\infty,I}
\nonumber\\
&+\frac{1}{|\alpha_{sp}|}\Big(L_{\log}\|\Delta\dot s\|_{\infty,I} +K_s\|\Delta s\|_{\infty,I}\Big).
\end{align}
\noindent\textbf{Step 3 (Stability of the inverse ODE for \(u\)).}

Let \(e:=u_1-u_2\). Using \eqref{eq:system_Psp_Q_sdot} gives
\[\dot e+a(s_1)e=\big(a(s_2)-a(s_1)\big)u_2+(F_1-F_2).\]
\[\text{Define } g(t):=\big(a(s_2)-a(s_1)\big)u_2+(F_1-F_2).\]
\begin{equation}
\label{eq:g_bound}
\text{Hence, }\|g\|_{\infty,I} \le L_aU\|\Delta s\|_{\infty,I}+\|F_1-F_2\|_{\infty,I}.
\end{equation}
Applying the standard variation-of-constants bound for
\[
\dot e+a(s_1)e=g
\]
on \(I\), and using \(a(s)\ge 0\), and no IC perturbation, and therefore
\begin{equation}
\label{eq:e_bound}
\|\Delta u\|_{\infty,I}\le T\|g\|_{\infty,I}.
\end{equation}
\noindent\textbf{Step 4 (Final constant).}

Finally, combining \eqref{eq:e_bound}, \eqref{eq:g_bound}, and \eqref{eq:F_diff}, together with \eqref{eq:delta_s_bound} and \eqref{eq:delta_sdot_bound}, yields
\[
\|u_1-u_2\|_{\infty,I}\leq C_H\|H_1-H_2\|_{W^{2,\infty}(I)}.
\]
An explicit, conservative bound is
\[
C_H:=T\Bigg[L_aU\,K_{s,H}+|b|+\frac{1}{|\alpha_{sp}|}\Big(L_{\log}K_{\dot s}+K_sK_{s,H}\Big)\Bigg],
\]
as stated in Theorem~\ref{thm:inverse_io_stability}.

% \subsubsection*{Effect of hyper-parameter margins on stability constant}
% \begin{itemize}
%     \item[(i)] The margin $\varepsilon_{sat}$ measures the distance from saturation of the sigmoid gating $s=T_p=\sigma(P-P_{sp})$. When $s\approx 0$ or $s\approx 1$, the reflex gating is effectively clamped, so variations in the neural set-point $P_{sp}(t)$ will only weakly affect the autonomic drives $T_p=s$ and $T_s=1-s$. Consequently, changes in $P_{sp}$ are only weakly reflected in the measured output $H$, making the inverse reconstruction more sensitive to noise and numerical errors. Thus, $\varepsilon_{sat}$ keeps the system in a physiologically responsive regime of the baroreflex.

%     \item[(ii)] The margin $\varepsilon_{deg}$ plays a non-degeneracy role in the inverse input--output mapping. Small values correspond to regimes where the measured dynamics $H$ and the gating $s$ provide little leverage on the set-point, so that the recovered $P_{sp}(t)$ becomes highly sensitive to perturbations. Larger values preserve a stronger coupling between output dynamics and regulatory drive, thereby improving the conditioning of the inverse reconstruction.

%     \item[(iii)] The window length $T$ controls the bias-variance trade-off in derivative recovery and therefore the gain by which noise in $H$ propagates to its derivatives, as quantified by $C_H$. A shorter window improves temporal resolution but amplifies noise, whereas a sufficiently long window reduces noise amplification and stabilizes estimation.
% \end{itemize}
In the theoretical expression of $C_H$, the dependence on $\varepsilon_{sat}$ arises through factors of the form $\big[\varepsilon_{sat}(1-\varepsilon_{sat})\big]^{-1}$, which explode as $\varepsilon_{sat}\to 0$. Hence, $C_H$ increases as $\varepsilon_{sat}$ decreases, while values closer to $0.5$ are less penalizing since $\varepsilon_{sat}(1-\varepsilon_{sat})$ is maximal at $\varepsilon_{sat}=0.5$.\\ By contrast, $\varepsilon_{deg}$ enters through inverse scalings such as $1/\varepsilon_{deg}$ and $1/\varepsilon_{deg}^2$, so increasing $\varepsilon_{deg}$ monotonically decreases $C_H$. Overall, the most favorable regime corresponds to sufficiently large $\varepsilon_{deg}$ together with moderate-to-large $\varepsilon_{sat}$.
\section{Joint unknown input and state estimation Physics-based Neural architecture}
\label{subsec:joint_estimator}
\begin{figure*}[ht]
  % \centering
  \hspace{0.5cm}
  \includegraphics[width=0.9\textwidth, trim={0 0 0 0.35cm}, clip]{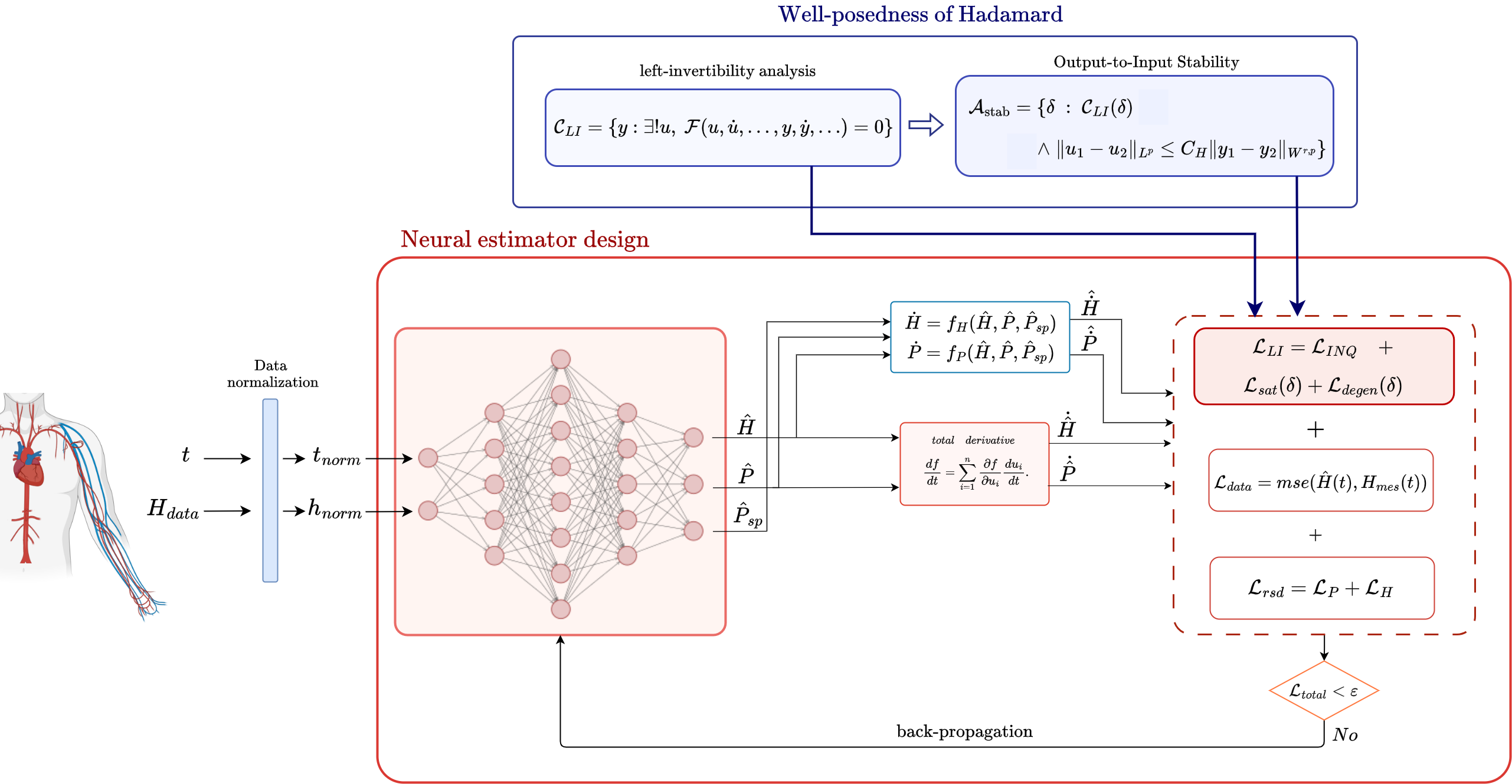}
  \caption{Architecture of the proposed Physics-Informed Neural Estimator for joint unknown control input $P_{sp}(t)$ and state $P(t)$ neural estimation (PIISNE)}
  \label{fig:pinns}
\end{figure*}

\noindent We propose a physics-based neural architecture that jointly estimates the unknown control input and the cardiovascular states under a coupled, nonlinear and controlled physiological model. In order to ensure well-posedness of the present inverse problem, uniqueness of the input reconstruction defined within a stability subspace; and enforce dynamical consistency, we inject physics-based residuals and left-invertibility constraints.\\
The estimator receives as inputs the normalized time variable \(t_{\mathrm{norm}}\) and a normalized, filtered signal \(h_{\mathrm{norm}}(t)\), derived from the measured heart rate \(H\). We normalize both inputs to avoid time acting as a diverging/scaling term when appearing in the estimates in the composite loss, and improve optimization conditioning and gradient flow. This architecture and estimation framework are illustrated in Figure \ref{fig:pinns}.
% \vspace{-0.8cm}
\noindent \subsection{\textbf{Loss formulation}}
Let \(\theta\) denote the network parameters and  $ \hat H(t, \theta), \hat P(t,\theta)$ are the estimated states and $\hat P_{\mathrm{sp}}(t, \theta)$ is the estimated control input.\\
Let a batch be
\(
\mathcal{B}=\{(t_{\mathrm{norm},i},\, h_{\mathrm{norm},i},\, H_{i})\}_{i=1}^{N}
\)
(normalized inputs; unnormalized measured data). The inverse problem is solved by minimizing the batch-averaged objective
\begin{equation}
\min_{\theta}\;\frac{1}{K}\sum_{b=1}^{K}\mathcal{L}_{\mathrm{total}}(\theta\mid\mathcal{B}_b).
\label{eq:empirical-risk}
\end{equation}
 where $K$ is the number of training batches and: $$\quad 
\mathcal{L}_{\mathrm{total}}
=\mathcal{L}_{\mathrm{data}}
+w_{\mathrm{phys}}\Big[
\mathcal{L}_{\mathrm{dyn}}+\mathcal{L}_{\mathrm{ID}}
\Big].
\label{eq:Ltotal}$$
\textbf{\textit{$(i)$ Data loss:}} $\mathcal{L}_{\mathrm{data}}$ accounts for data fidelity loss, $\mathcal{L}_{\mathrm{dyn}}$ for the model residuals and $\mathcal{L}_{\mathrm{ID}}$ for the identifiability constraints.\\
\noindent Data and physics terms are balanced through an adaptive factor:
$\qquad
w_{\mathrm{phys}}
=
\frac{\mathcal{L}_{\mathrm{data}}}
{\mathcal{L}_{\mathrm{dyn}}+\mathcal{L}_{\mathrm{ID}}},
\label{eq:wphys}$

For the experimental data learning, we compare the estimate to the unnormalized measurement to preserve physical units:
\begin{equation}
\mathcal{L}_{\mathrm{data}}(\theta\mid\mathcal{B})
=\sum_{i\in\mathcal{B}}\!\big\|\hat H(t_i,\theta)-H_{i}\big\|_2^2.
\label{eq:Ldata}
\end{equation}
\textbf{\textit{$(ii) $ Model's residuals' loss:}} $\mathcal{L}_{\mathrm{dyn}}$ enforces the agreement between the temporal evolution predicted by the neural estimator and the physiological  yielded by the dynamical system:
\begin{equation}
\begin{aligned}
\mathcal{L}_{\mathrm{dyn}}(\theta\mid\mathcal{B})
&=
\sum_{i\in\mathcal{B}}\Big\|
\dot{\hat P}(t_i,\theta)-f_P\!\big(\hat{x}_\theta(t_i),\hat{u}_\theta(t_i)\big)
\Big\|_2^2 \\
&\quad+
\sum_{i\in\mathcal{B}}\Big\|
\dot{\hat H}(t_i,\theta)-f_H\!\big(\hat{x}_\theta(t_i),\hat{u}_\theta(t_i)\big)
\Big\|_2^2.
\end{aligned}
\label{eq:Ldyn}
\end{equation}
\noindent \textbf{\textit{Remark:}} A crucial difference of the present work with respect to frameworks aimed at solving inverse problems \cite{raissi2019physics, kim2024review} resides in the reconstruction of the control input, which depends not only on time but, more importantly, on the time-varying dynamics it induces in the measured output. This naturally leads to the necessity of computing the total derivatives rather than relying only on the partial derivative w.r.t. time auto-differentiation outputs, which is the standard approach in PINNs. We capture both explicit temporal dependence and implicit dependence through the evolving $h_{norm}$:
\begin{equation}
\begin{cases}
\dfrac{d\hat H}{dt}=\dfrac{\partial \hat H}{\partial t}
+\dfrac{\partial \hat H}{\partial h_{\mathrm{norm}}}\dfrac{dh_{\mathrm{norm}}}{dt},\\[8pt]
\dfrac{d\hat P}{dt}=\dfrac{\partial \hat P}{\partial t}
+\dfrac{\partial \hat P}{\partial h_{\mathrm{norm}}}\dfrac{dh_{\mathrm{norm}}}{dt}.
\end{cases}
\label{eq:total-derivs}
\end{equation}
Relying only on automatic differentiation with respect to \(t\) would omit the \(\partial(\cdot)/\partial h_{\mathrm{norm}}\) pathways and thus fail to represent the temporal evolution of the dynamics captured by the physiological model.\\
No explicit dependence of the measurement data with respect to time is available. Therefore, the derivative of $h_{norm}$ w.r.t time is computed using simple numerical differentiation (finite differences) applied to the filtered signal to avoid noise amplification.

\textbf{\textit{Remark:}} Although time derivatives of the measured output are not directly accessible in practice, they can be approximated from sampled trajectories using numerical or algebraic differentiation schemes \cite{liu2011error}. While the left-invertibility conditions are stated for continuous-time signals over a time window and assume access to the required derivatives, the estimator adopted in this work operates on discrete measurements: the necessary derivatives are obtained numerically, and we show that with appropriate smoothing and hyper-parameter selection, the associated noise amplification can be controlled.

\noindent \textbf{$(iii)$ \textit{Left-invertibility loss:}} We inject the structural identifiability constraints derived above through
\begin{equation}
\mathcal{L}_{\mathrm{ID}}
=
\mathcal{L}_{\mathrm{INQ}}
+
\mathcal{L}_{\mathrm{sat}}
+
\mathcal{L}_{\mathrm{degen}}.
\label{eq:LID}
\end{equation}
Violations of the admissible interval are penalized with$$ \mathcal{L}_{\mathrm{INQ}}
= \frac{1}{N}\sum_{k=1}^N
\operatorname{softplus}(-V_H-\hat Y_k)
+ \frac{1}{N}\sum_{k=1}^N
\operatorname{softplus}(\hat Y_k-\beta_H),$$
prevent saturation of the sigmoid by keeping $\hat s_k$ through $$ \mathcal{L}_{\mathrm{sat}}
=
\lambda_{\mathrm{sat}}\,\frac{1}{N}\sum_{k=1}^N
\Bigl[\operatorname{softplus}\bigl(\hat s_k-\text{logit}(\varepsilon_{sat}) \bigr)\Bigr]^2,$$
and non-degeneracy of \eqref{eq:Psp_dae} with a margin from zero $$
\mathcal{L}_{\mathrm{degen}}
=\frac{1}{N}\sum_{k=1}^N
\Bigl[\operatorname{softplus}\bigl(\varepsilon_{deg}-|D|\bigr)\Bigr]^2.$$
Here, $\varepsilon_{\mathrm{sat}}>0$ and $\varepsilon_{\mathrm{degen}}>0$ are the margin hyper-parameters tuned through estimation stability analysis.

We recall the need for a boundary condition on $P$ to ensure input identifiability. In the present work, we set an inequality boundary condition as $P\geq 60$ to ensure physiologically-coherent estimation.
\section{Stability constant: numerical simulations}
We perform numerical simulations of $C_H(\varepsilon_{sat},\varepsilon_{deg})$ to quantify how sensitive this stability constant is to the margin hyperparameters $\varepsilon_{sat}$ and $\varepsilon_{deg}$, and hence to assess the tightness and conditioning of the associated bound. In our setting, a smaller $C_H$ implies that output-to-input error amplification is controlled, guaranteeing that convergence of the estimated output toward the target entails a comparably bounded convergence of the reconstructed input. The goal is therefore to delineate safe regions where $C_H$ remains small-to-moderate, guaranteeing stability of the inverse Input-Output mapping.

\textbf{Simulation setup.}
We evaluate $C_H$ on a grid with $\varepsilon_{sat}\in[10^{-3},\,0.49]$ and $\varepsilon_{deg}\in[10^{-3},\,10]$, using a duration parameter equivalent to $40$ samples, consistent with that used later for smoothing excessive noise in numerical derivatives. We fix the rest of the parameters, including the upper bounds of the states and intermediate variables, to physiologically coherent values to limit the degrees of freedom to the hyperparameters of interest, and proceed with an uncertainty analysis on the remaining fixed parameters.

\begin{figure*}[t]
    \centering
    \subfloat[Sensitivity surface $\log_{10}(C_H)=f(\varepsilon_{sat},\varepsilon_{deg})$.%
    \label{fig:CH_3D_surface}]{
        \includegraphics[width=0.44\linewidth,trim={0 0 0 1.9cm},clip]{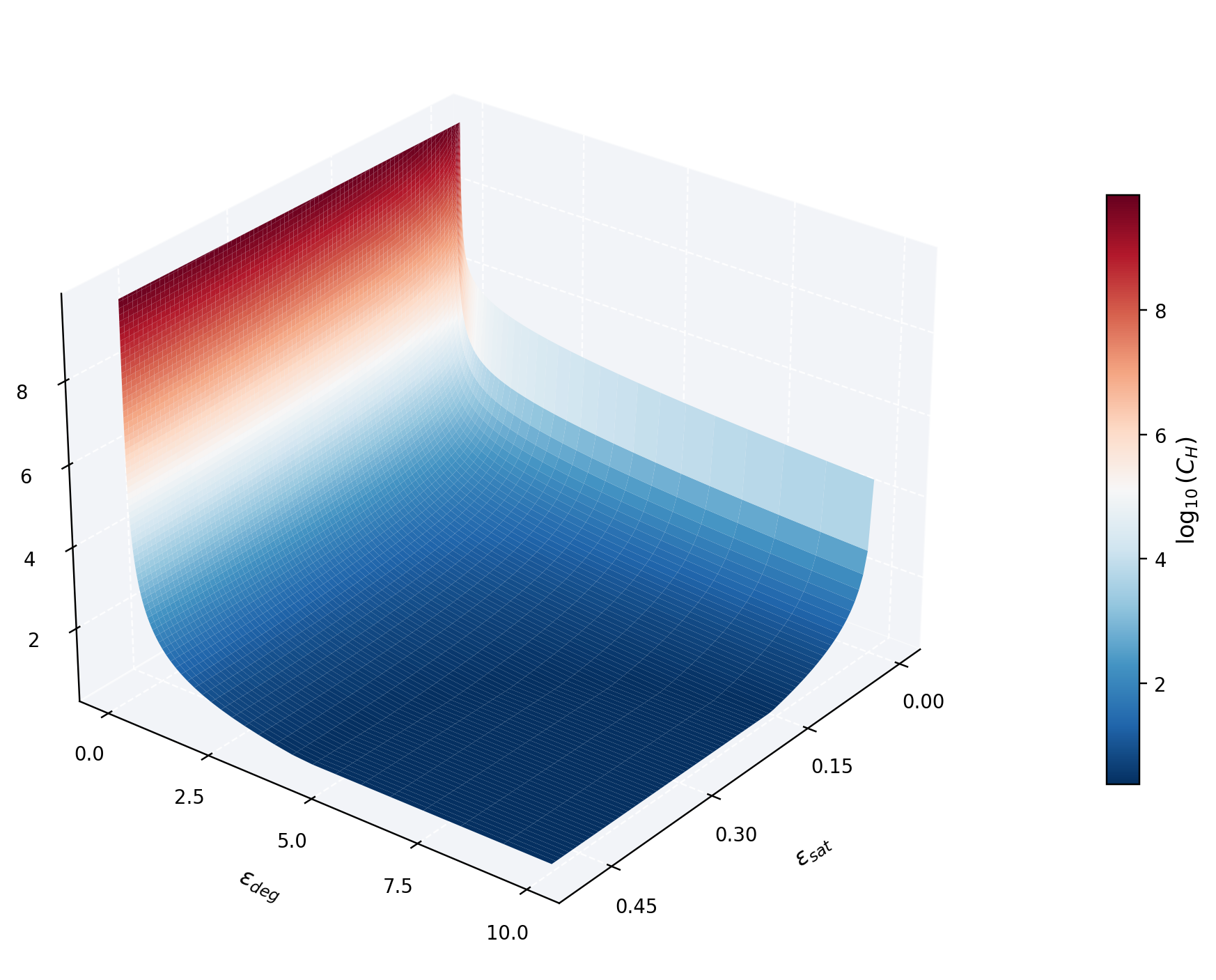}
    }
    \subfloat[Uncertainty with respect to the uncertain parameters $\{M_{\dot Y}, M_{\dot S}\}$.%
    \label{fig:uncertainty_CH}]{
        \includegraphics[width=0.54\linewidth,trim={0 0 0 1cm},clip]{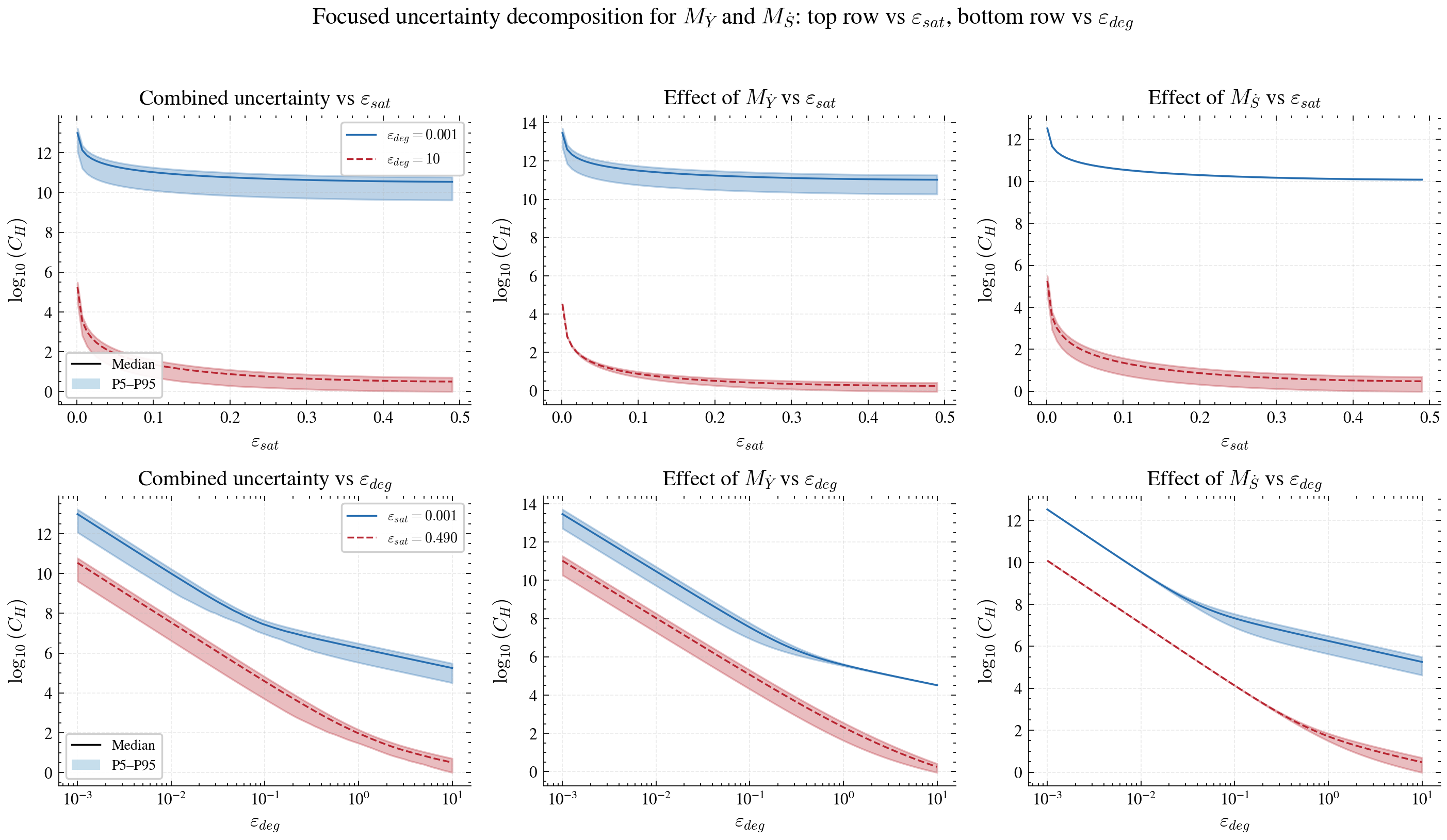}
    }
    \caption{Sensitivity analysis of the Lipschitz constant of the estimator with respect to the margin and uncertain parameters.}
    \label{fig:CH_sensitivity}
\end{figure*}

Figure \ref{fig:CH_3D_surface} reports the 3D response surface of \\ $\log_{10}\!\big(C_H(\varepsilon_{sat},\varepsilon_{deg})\big)$ over the $(\varepsilon_{sat},\varepsilon_{deg})$ plane. In addition, a correlation-based sensitivity analysis was carried out with respect to the remaining fixed bounds to identify the parameters exerting the strongest influence on $C_H$. The Monte Carlo uncertainty analysis shown in Figure \ref{fig:uncertainty_CH} was then performed only on these selected sensitive quantities. Taken together, Figure~\ref{fig:CH_sensitivity} provides a compact view of how sensitivity and uncertainty shape the stability bound across the parameter domain.\\
The numerical sweep in Figure \ref{fig:CH_3D_surface} is consistent with the foregoing analysis, with increasing $\varepsilon_{deg}$ producing a monotone decrease in $C_H$. While the dominant source of variability in $C_H$ is $\varepsilon_{deg}$, the influence of $\varepsilon_{sat}$ remains visible: moving away from the lower-bound region, i.e. as $\varepsilon_{sat}$ increases toward $0.5$, decreases $C_H$, although it progressively saturates as $\varepsilon_{sat}$ reaches moderate values.\\
This transition from a poorly conditioned to a more stable regime is rapid near the smallest admissible values and progressively slower as both hyperparameters increase. In this interpretation, $\varepsilon_{deg}$ primarily controls the regularization strength induced by the left-invertibility constraints, whereas $\varepsilon_{sat}$ determines the admissible tracking margin between the blood pressure state $P$ and the estimated input trajectory $P_{sp}$. Hence, the smallest values of $C_H$ are obtained for sufficiently large $\varepsilon_{deg}$ combined with moderate-to-large $\varepsilon_{sat}$. These simulations are therefore used to guide the selection of admissible hyperparameter values according to the tolerance level the practitioner is willing to impose, rather than through ad hoc fixing or blind tuning. In particular, an operational regime can be identified in which $\log_{10}(C_H)_{\min}\approx -0.08$, suggesting approximate equivalence between the estimation errors of $H$ and $P_{sp}$, with the possibility of mild denoising.\\
The Monte Carlo uncertainty analysis results displayed in Figure \ref{fig:uncertainty_CH} confirm that the deterministic dependence of $C_H$ on $\varepsilon_{sat}$ and $\varepsilon_{deg}$ remains valid under uncertainty propagation in the rest of the parameters entering the bound, as the median curves preserve the same ordering as in the nominal study, with a markedly stronger effect for $\varepsilon_{deg}$. The percentile envelopes further show that, for $\varepsilon_{sat}$, the envelope width decreases mainly near its lower-bound region and then rapidly saturates, consistent with its role in preventing saturation of the gating variable $S$. This effect is more pronounced for the $M_{\dot S}$-related term, while the $M_{\dot Y}$-related contribution remains a more persistent source of dispersion due to its sensitivity to output derivatives and noise amplification. By contrast, increasing $\varepsilon_{deg}$ yields a stronger and more sustained reduction of both the median and the envelope width, indicating that it acts as the main global conditioning parameter of the inverse mapping. Notably, the envelopes tend to extend further below than above the median, an asymmetry that is favorable given that smaller values of $C_H$ imply better conditioning.
\vspace{-0.3cm}
\section{Data}
\subsection{The PASS dataset}
\noindent The PASS dataset \cite{parent2020pass}  proposes a bank of multiple signals collected from $48$ participants asked to perform tasks of varying levels of emotional and physical stress and wearable sensors were used to record cardiac,  respiratory activity, and cerebral activities.\\
This study relies on Heart-Rate time-series extracted from mobile ECG recordings. We first downsample all signals to $5$ $Hz$ to reduce computational load and deliberately operate in a temporally sparse regime. This down-sampling preserves characteristic slow autonomic modulations $(<0.5Hz)$. Yet, the short-term autonomic response targeted presents faster dynamics.\\
% Figure \ref{fig:KDE_data} illustrates a ridge distribution of the temporal data per subject in the dataset.\\
% \begin{figure}[htbp]
%     \centering
%     \includegraphics[width=0.5\linewidth, trim={0 0 0 1.6cm}, clip]{images/ridgeline_KDE_data_RdBu.png}
%     \caption{Ridge distribution per participant key}
%     \label{fig:KDE_data}
% \end{figure}
From the time-series of each subject-specific (key), we train a dedicated model using $7500$ samples extracted as contiguous segments, so as to preserve temporal coherence and dependencies. The remaining samples from the same participant constitute the held-out test set for the within-subject evaluation. For cross-subject evaluation, the model trained on the time-series of a given participant is evaluated on the full time series of each unseen target key, i.e. on participants entirely excluded from that model’s training.
\vspace{-0.2cm}
\subsection{the 3CI dataset}
The present dataset is multimodal physiological database acquired in $30$ participants for the study of brain–heart interaction during acute stress. It includes simultaneous $32$-electrodes EEG, ECG, and continuous beat-to-beat non-invasive arterial blood pressure recorded with Finometer. Three validated stress paradigms were included for social-evaluative stress, nociceptive and sympathetic activation, and a graded physical effort, each accompanied by resting baseline and recovery periods. Cross-modal synchronization was ensured to support precise temporal alignment of recordings. The present dataset is used exclusively for testing and validating upon both variables $H$ and $P$.
\vspace{-0.2cm}
\subsection{Simulated data}
The simulated dataset comprises 500 synthetic envelope signals representing $P_{sp}$. Each envelope is generated over $1000$ s at an initial sampling step of $0.005$ s, starting from a randomly selected baseline and including three successive bump-shaped modulations with variable rise times, plateau durations, decay phases, and amplitudes. This construction introduces variability while preserving a common temporal structure coherent with autonomic task-specific responses. After generation, the envelopes are downsampled by a factor of $20$. These $500$ $P_{sp}$ profiles are then used as inputs to the dynamical system to simulate the corresponding trajectories of $H$ and $P$, thereby producing a complete synthetic dataset. Table \ref{tab:datasetss} recalls available data for each dataset.
\begin{table}[htbp]
    \centering
    \renewcommand{\arraystretch}{1.5}
    \setlength{\tabcolsep}{15.5pt} % increase column width
    \begin{tabular}{|c|c|c c c|}
    \cline{3-5}
    \multicolumn{2}{c|}{} & $H$ & $P$ & $P_{sp}$ \\
    \hline
    \multirow{3}{*}{\rotatebox{90}{DATASET}}
    & PASS      & $\checkmark$ & \ding{55} & \ding{55} \\
    \cline{2-5}
    & Simulated & $\checkmark$ & $\checkmark$ & $\checkmark$ \\
    \cline{2-5}
    & 3CI       & $\checkmark$ & $\checkmark$ & \ding{55} \\
    \hline
    \end{tabular}
    \vspace{0.3cm}
    \caption{Summary of available recordings and variables for each dataset.}
    \label{tab:datasetss}
\end{table}
\vspace{-0.3cm}
\section{Simulations and Results}
\subsection{Neural Network architecture and training scheme}
\noindent The neural component is a feed-forward multilayer perceptron (MLP) with seven fully-connected layers \([2,\,10,\,25,\,40,\,20,\,10,\,3]\) and \textsf{GELU} activations; such activation functions were chosen for its smooth, twice-differentiable nonlinearity, which supports stable physics-based derivative constraints. We do not include an activation function in the last layer for freedom of output values. Given \([t,\,h_{\mathrm{norm}}(t)]\), the model produces the estimated system states \(\hat{x}_\theta(t)\) and reconstructed input \(\hat{u}_\theta(t)\).\\
Although \(H\) is directly observable, the network re-estimates \(\hat H(t, \theta)\) from \([t,\,h_{\mathrm{norm}}(t)]\) in order to preserve temporal coupling with the other estimated states, enable residual-based physics enforcement, and ensure gradients propagate consistently through all dynamical dependencies.
% The physiological model includes a known, constant \emph{state delay} \(\tau\), represented through the delayed pressure \(\hat P_\theta(t-\tau)\). We implement this term via a differentiable history buffer (method-of-steps) that stores recent values of \(\hat P_\theta\) and retrieves the delayed state during training while preserving backpropagation.

\textbf{Training and stopping criteria:} Training proceeds end-to-end in PyTorch using Adam (learning rate \(0.1\), batch size \(1500\). Training is ran for a maximum of $95$ epochs with early stopping based on validation performance. The checkpoint is updated only when the improvement exceeds $10^{-2}$ relative to the current best score. If no such improvement is observed for $20$ consecutive epochs, training is terminated and the best-performing checkpoint within the preceding window is retained as the final set of weights.\\
We use Python $3.14.2$ software for the simulations on a PC with $32$ GB RAM, Intel Core $i7-13800H$ CPU on Windows $10$ operating system, and a $\text{NVIDIA RTX Ada}$ GPU.

\noindent \textbf{\textit{Remark:}} The omission of the state time-delay in the present dynamical system constrains the variability of the states. In this scope, a regularization term is added to encourage it for the estimates:
\begin{equation}
\mathcal{L}^{'}_{total}=\mathcal{L}_{total}-\lambda\, \text{std}\big[\text{logit}(s)\big]
\end{equation}
\vspace{-0.7cm}
\subsection{Simulation results and discussion}
\begin{figure*}[t]
    \centering
    \includegraphics[width=0.95\linewidth, trim={0 0.13cm 0 0.57cm}, clip]{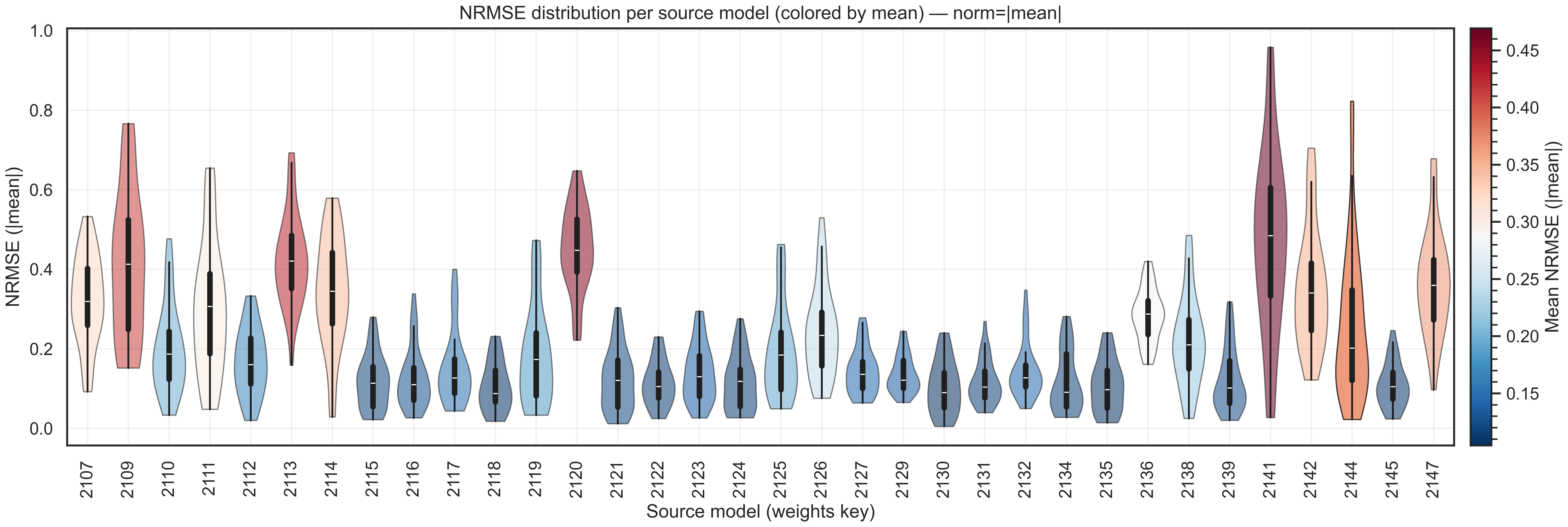}
    \caption{NRMSE distribution per model.}
    \label{fig:nrmse_violins}
\end{figure*}

\begin{figure*}[t]
    \centering
    \begin{minipage}[t]{0.51\linewidth}
        \vspace{0pt}
        \centering
        \subfloat[Heatmap of estimation performance (NRMSE) for each model trained on one key (columns) and evaluated on all keys (rows).%
        \label{fig:heatmap_left}]{
            \includegraphics[width=\linewidth,trim={0 0 0 0.58cm},clip]{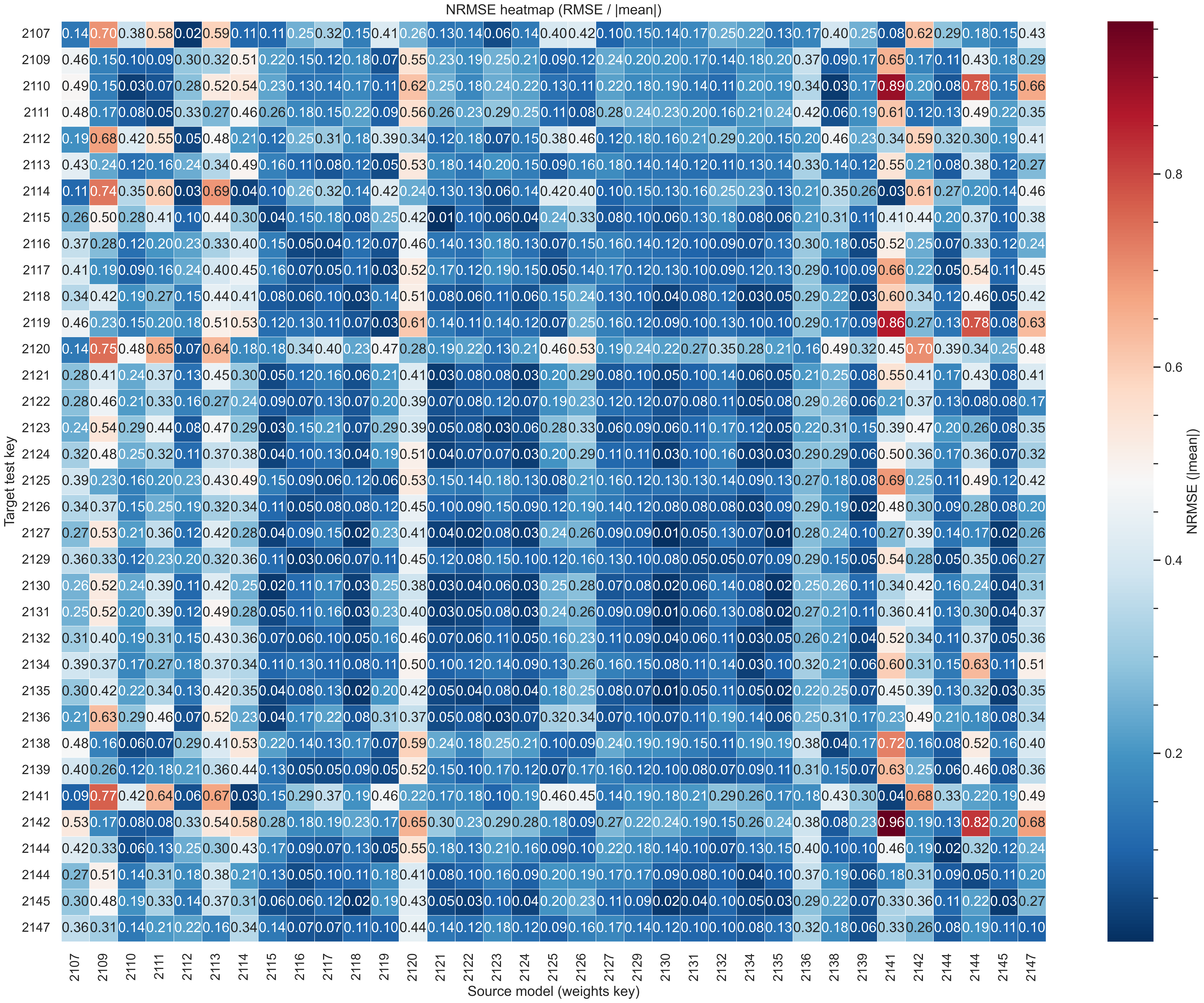}
        }
    \end{minipage}
    \hfill
    \begin{minipage}[t]{0.48\linewidth}
        \vspace{0.2cm}
        \centering
        \subfloat[Best off-diagonal configuration.%
        \label{fig:best_offdiag}]{
            \includegraphics[width=\linewidth]{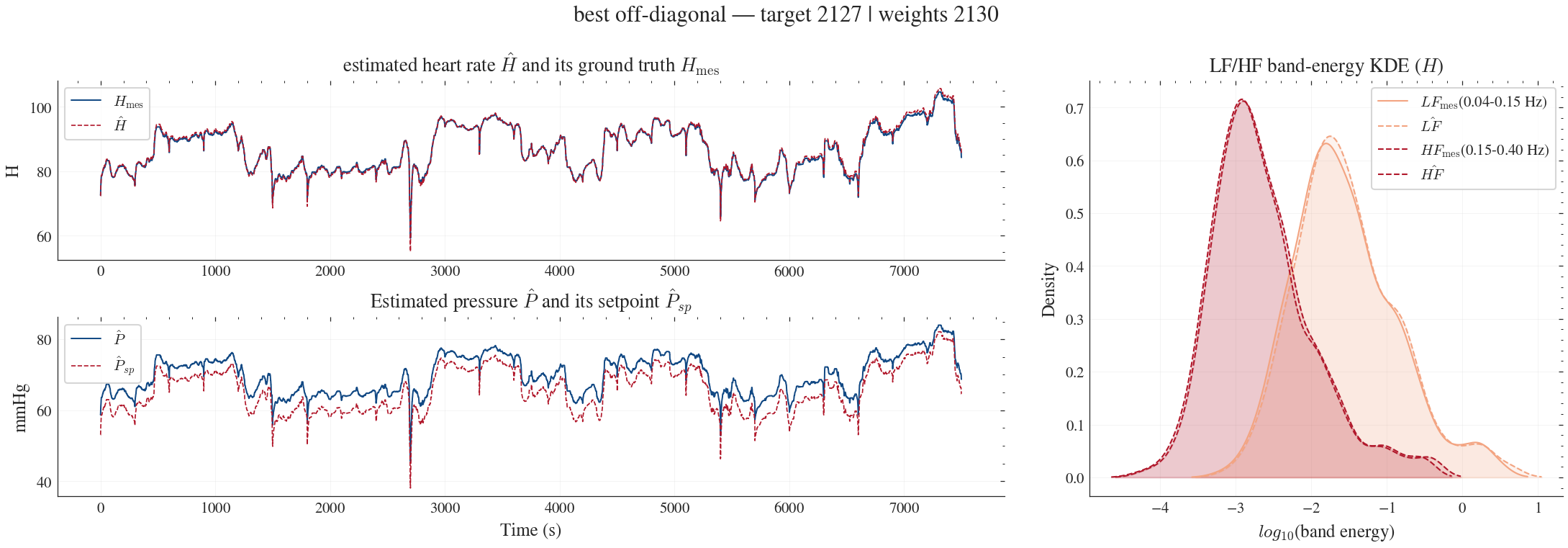}
        }

        \vspace{0.35cm}

        \subfloat[Worst off-diagonal configuration.%
        \label{fig:worst_offdiag}]{
            \includegraphics[width=\linewidth]{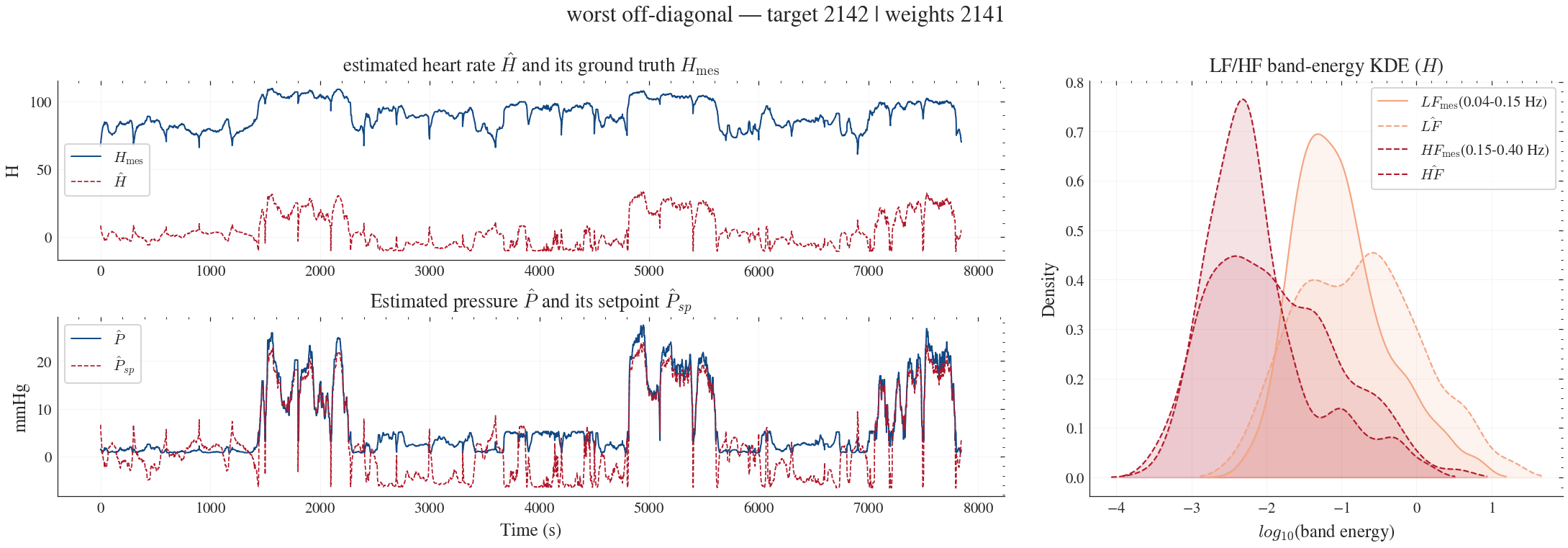}
        }
    \end{minipage}
    \caption{Overall comparison of estimation performance: heatmap of cross-key results (left), and representative best and worst off-diagonal configurations (right).}
    \label{fig:cross_sub_perf}
\end{figure*}

\noindent Figures \ref{fig:nrmse_violins} and \ref{fig:cross_sub_perf} summarize the cross-participant behavior of the estimator when the neural-network weights trained on one participant (source key) are deployed, without recalibration or adaptation, to reconstruct the measured hemodynamic output of the rest of subjects (target keys). Figure \ref{fig:nrmse_violins} shows that several source keys yield narrow distributions centered at low NRMSE, demonstrating that the PIISNE can generalize to previously unseen participant data while jointly inferring the cardiovascular variables ($\hat P$ and $\hat H$) and the corresponding unknown control input (interoceptive $\hat{P}_{sp}$). This is particularly visible for sources in the $2115$--$2139$ range, whose narrow, low-centered violin plots confirm robustness to inter-subject variability rather than source-specific fitting. By contrast, a smaller number of sources exhibit larger errors together with broader distributions, indicating parameterizations that are less transferable.\\
\noindent Figure \ref{fig:heatmap_left} refines this observation by exposing the structure of the cross-participant error map. The observed degradation patterns are consistent with deployments in which the observable does not fully constrain the latent dynamics, or with spectral bias in the network.\\
\noindent The best and worst off-diagonal cases identified in Figure \ref{fig:heatmap_left} are illustrated in Figures \ref{fig:best_offdiag} and \ref{fig:worst_offdiag}. In the best off-diagonal configuration (target $2127$ evaluated with weights $2130$; Figure \ref{fig:best_offdiag}), the reconstructed $\hat{H}$ closely follows $H_{\mathrm{data}}$ over both slow trends and faster autonomic fluctuations, with limited bias over the full concatenated horizon, while the LF and HF band-energy KDEs closely overlap those of the measured signal, indicating that the transferred weights preserve not only time-domain reconstruction but also the oscillatory content in the canonical autonomic bands. While both $P(t)$ and the neural setpoint $P_{sp}(t)$ are inferred without direct targets, the strong proximity observed between $\hat{P}(t)$ and $\hat{P}_{sp}(t)$ is physiologically expected and coherent: $P_{sp}(t)$ represents the neural setpoint that blood pressure $P(t)$ must dynamically track, consistent with the comparator role of the Nucleus Tractus Solitarius in integrating hemodynamic signals with descending neural control.\\
\noindent By contrast, the worst off-diagonal case (target $2142$ evaluated with weights $2141$; Figure \ref{fig:worst_offdiag}) shows a qualitatively different regime. The reconstructed $\hat{H}$ collapses toward a lower-amplitude trajectory, indicating that the deployed parameterization is not compatible with the target dynamics under the same constrained inference. This mismatch propagates to the inferred latent pressures: both $\hat{P}$ and $\hat{P}_{sp}$ drift toward implausibly low values and exhibit intermittent burst-like excursions rather than smooth coupled modulation. The LF/HF distributions are likewise degraded, confirming a failure of cross-subject transfer in which the constrained learned mapping cannot reconcile the target's baseline and responsiveness with the source parameterization.\\
\noindent To assess generalization more rigorously, the previously trained PIISNE was also evaluated on both the simulated and 3CI datasets. This offers a clearer assessment of its generalization capability and allows validation on variables ($P$, $P_{sp}$) that were not available in the PASS dataset. The performance of the best-identified source key across all three datasets is reported in Table \ref{tab:sum_perf}.\\
\begin{table*}[t]
    \centering
    \renewcommand{\arraystretch}{1.5}
    \begin{tabular}{|c|c|c|c|c|c|}
        \cline{3-6}
        \multicolumn{2}{c|}{} & src key & $H$ & $P$ & $P_{sp}$ \\
        \hline
        \multirow{3}{*}{\rotatebox{90}{DATASET}}
        & PASS      & $2130$ & $0.099 \pm 0.07$ & \cellcolor{gray!40}{ } & \cellcolor{gray!40}{ } \\
        \cline{2-6}
        & Simulated & $2130$ & $0.04 \pm 0.02$  & $0.046 \pm 0.03$ & $0.1 \pm 0.05$ \\
        \cline{2-6}
        & 3CI       & $2126$ & $0.16 \pm 0.07$  & $0.21 \pm 0.08$ & \cellcolor{gray!40}{ } \\
        \hline
    \end{tabular}
    \caption{Summary of the performance of the best-performing neural network, trained on the PASS source key, and its generalization across participants from the simulated, PASS, and 3CI datasets, reported as symmetric mean absolute percentage error (SMAPE).}
    \label{tab:sum_perf}
\end{table*}
\noindent As expected, the lowest errors are obtained on the simulated dataset, with larger errors on the two real datasets reflecting their greater variability and heterogeneous recording conditions. Yet, these errors display substantially small standard deviations, indicating stability in the estimation. This crucially validates that the estimator is capable of jointly reconstructing unknown control input and state from a single observable. Closer inspection reveals that the main discrepancy is an offset, while the estimator still captures the underlying dynamics reasonably well, likely attributable to the use of fixed physiological parameters — in particular the stroke volume parameter, which is highly sensitive.\\
\noindent Across variables, errors increase from simulated to real data for both $H$ and $P$, the latter plausibly reflecting baroreflex regime shifts induced by task diversity in the 3CI protocol. The estimation of $P_{sp}$, available only in the simulated dataset, remains highly accurate, providing a direct validation of the proposed scheme for joint state and control-input inference from a single observable.
\vspace{-0.3cm}
\section{Robustness and comparison}
\vspace{-0.2cm}
\subsection{Ablation study}
\begin{figure*}[htbp]
    \centering
    \subfloat[2115\label{fig:concat_2115}]{
        \includegraphics[width=0.33\linewidth,trim={0 0 0 0.3cm},clip]{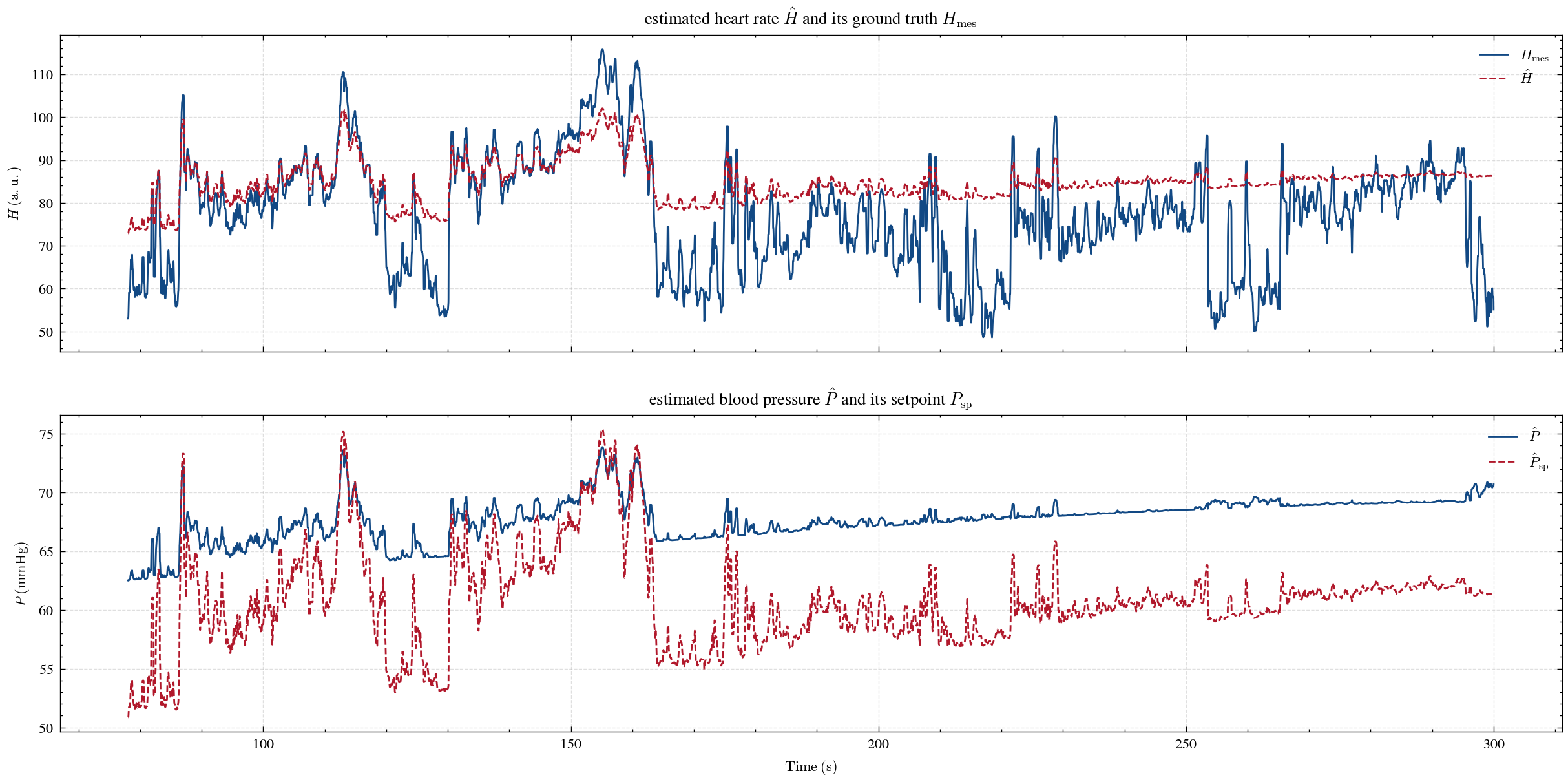}
    }
    \subfloat[2117\label{fig:concat_2117}]{
        \includegraphics[width=0.33\linewidth,trim={0 0 0 0.3cm},clip]{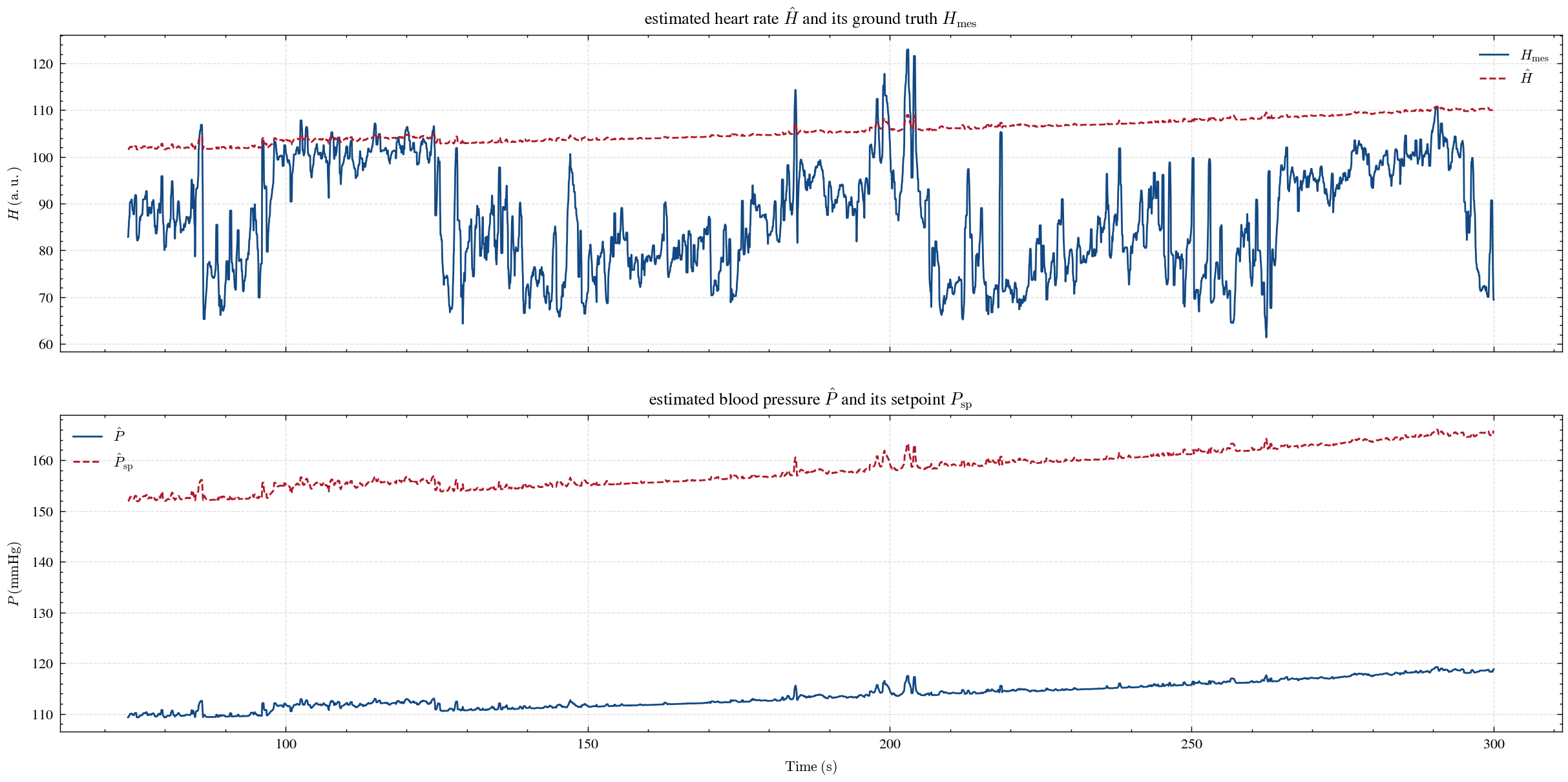}
    }
    \subfloat[2121\label{fig:concat_2121}]{
        \includegraphics[width=0.3\linewidth,trim={0 0 0 0.3cm},clip]{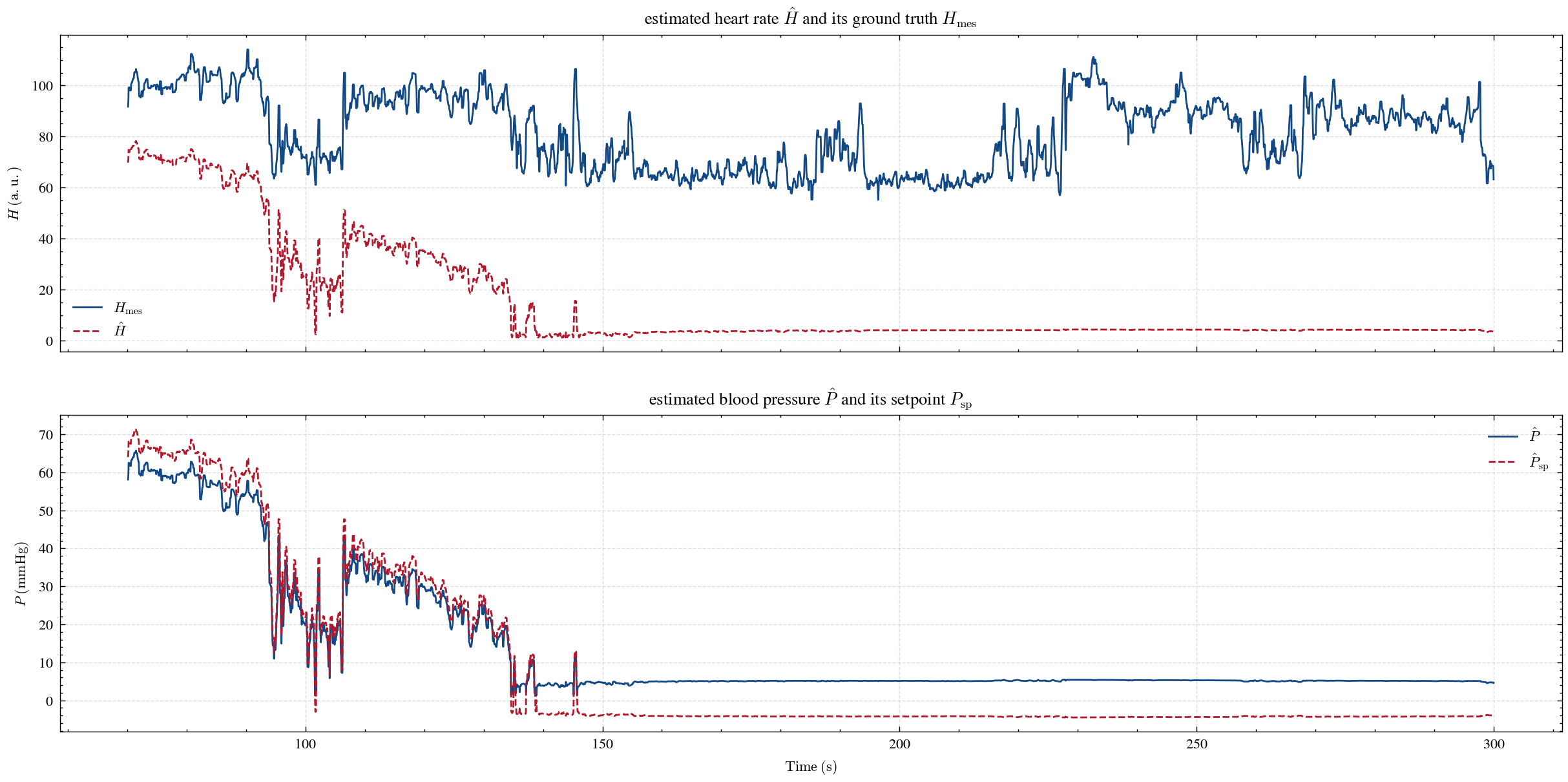}
    }
    \caption{Illustrative failure cases after ablation of the left-invertibility loss terms.}
    \label{fig:ablation}
\end{figure*}

This section presents an ablation study designed to assess the contribution of the main components of the proposed framework. By selectively removing exclusively the well-posedness components from the loss, we evaluate their impact on the overall performance and robustness of the model. This analysis provides a clearer assessment of its contribution to the performance of the neural estimator. We focus the testing of the models on never-seen samples from the same subject (key) it was trained on -no generalization requirement. Figure \ref{fig:ablation} provides three illustrations of failure cases for this self-prediction after ablation.\\
All three cases are striking illustrations as all performed highly on self-prediction as seen in Figure \ref{fig:heatmap_left}, and tangibly showcase the need and contribution to incorporate well-posedness in the learning design. Indeed, the estimation of heart rate collapses and the other estimates are non-physiological and collapse similarly, coherent with the functioning of controlled autonomic activity.
\vspace{-0.3cm}
\subsection{Model-based estimator: Sliding-mode observer}
\begin{figure*}[htbp]
    \centering
    \subfloat[2112\label{fig:2112_smo}]{
        \includegraphics[width=0.31\linewidth,trim={0 0 0 1cm},clip]{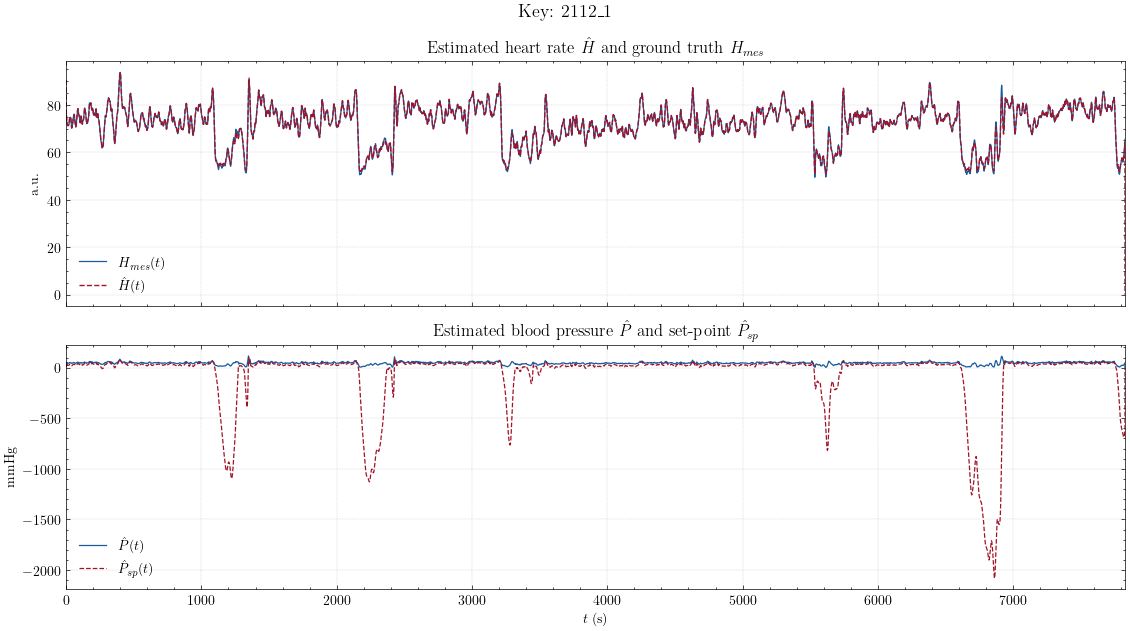}
    }
    \subfloat[2115\label{fig:2115_smo}]{
        \includegraphics[width=0.31\linewidth,trim={0 0 0 1cm},clip]{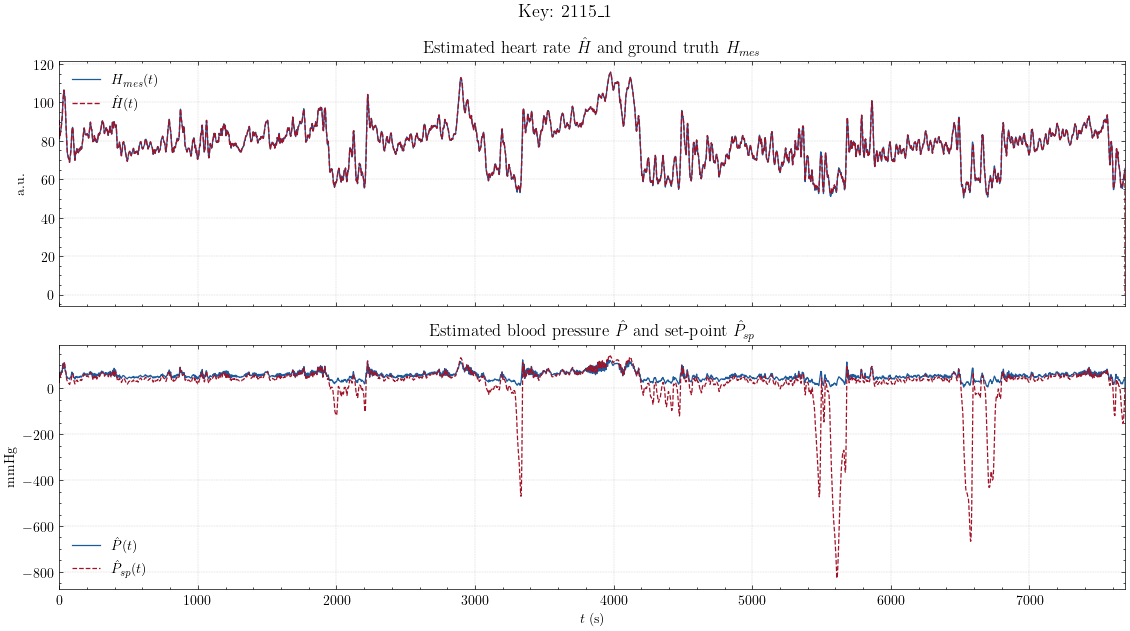}
    }
    \subfloat[2134\_1\label{fig:2134_smo}]{
        \includegraphics[width=0.31\linewidth,trim={0 0 0 1cm},clip]{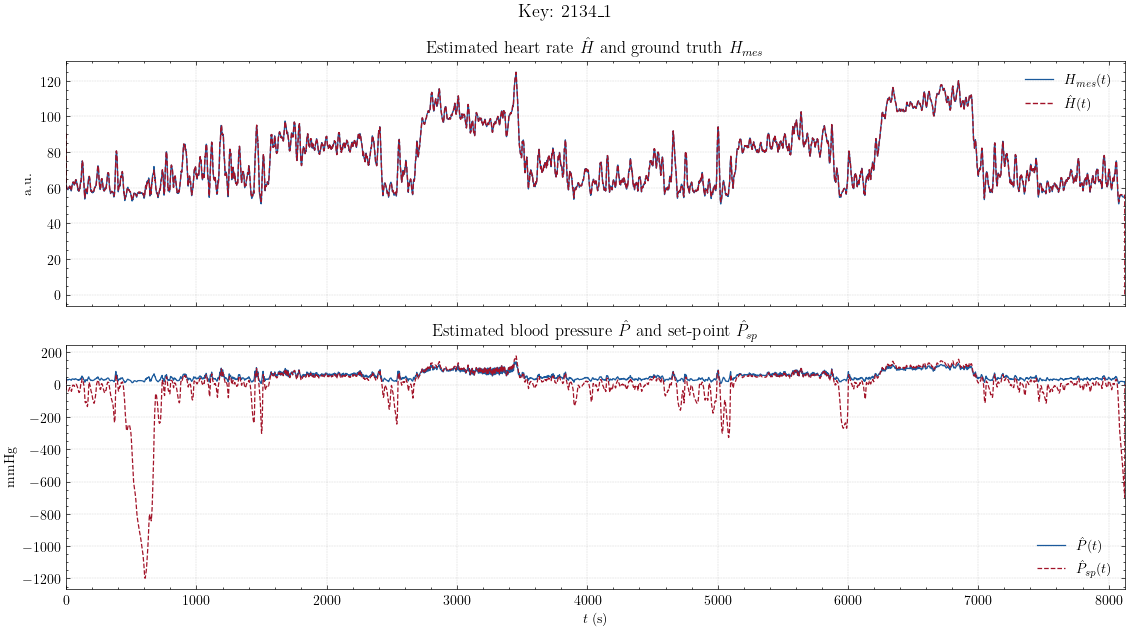}
    }
    \caption{Illustrative cases of \(P_{sp}\) reconstruction failure despite measurement fidelity.}
    \label{fig:SMO}
\end{figure*}

As a model-based baseline for comparison with the neural estimator, we consider a sliding-mode observer (SMO) as a model-based estimator of the latent set-point $P_{sp}$. The formulation assumes that $P_{sp}$ behaves as a slowly varying augmented state over the estimation window, a dynamical assumption that may, however, introduce model mismatch. The SMO is nonetheless well suited for nonlinear estimation from limited measurements and exhibits inherent robustness to such mismatch, which motivates its selection as a comparison point. The observer gains are calibrated jointly over all recordings through bounded global optimization, by minimizing the average normalized reconstruction error on the measured heart-rate. The resulting estimate therefore corresponds to one common observer parameterization, selected for cross-recording performance rather than per-trajectory fitting. Figure \ref{fig:SMO} provides three illustrative cases of fidelity to $H$ coexisting with failure in $P_{sp}$ estimation.\\
These illustrative failures showcase poor, non-physiological recovery of $P_{sp}$ coexisting with output fidelity on $H$: aberrant chattering-like transients in the estimation of $P_{sp}$, a behavior well-documented in SMO-based estimation. These excursions also reflect a limitation of the augmented-state formulation itself, as $P_{sp}$ is reconstructed under a slow-variation assumption that may further degrade the estimate and promote non-physiological dynamics. These results illustrate the core limitation of purely model-based estimators in the present problem formulation: their inherent dependency on assumptions regarding the dynamics of the unknown input to be estimated.
\vspace{-0.35cm}
\section{Conclusion}
\label{sec:CONCLUSIONS}
\noindent As estimation problems are increasingly recast as optimization problems to leverage neural networks as universal function approximators, the classical control-theoretic requirements attached to estimation can no longer remain implicit. In Scientific Machine Learning (SciML), this translation is often implemented through residual terms associated with the governing dynamics; in inverse inference, however, the central issue is not merely trajectory fitting but whether unknown-input reconstruction is well posed in the Hadamard sense \cite{kabanikhin2008definitions}. In this perspective, rather than relying only on generic regularization to bias learning toward plausible solutions, we adopted a complementary paradigm in which control-theoretic solvability requirements directly inform both estimator design and training.\\
\noindent Within this framework, we proposed a physics-informed neural estimator for the joint reconstruction of blood pressure, heart rate, and the latent neural control input. For the considered controlled dynamical system, left-invertibility provides the relevant structural notion of uniqueness, and we showed how the associated conditions can be transferred into the learning formulation through dedicated loss terms. Beyond uniqueness, Output-Input stability was addressed through the derivation of an explicit Lipschitz-like constant, whose numerical analysis clarified how the left-invertibility margins shape the conditioning and robustness of the inverse problem. The associated deterministic and uncertainty analyses consistently identified the most favorable hyperparameter regime for stable reconstruction.\\
\noindent The numerical results demonstrate accurate reconstruction of cardiovascular dynamics while enabling recovery of the latent regulatory input, interpreted here as an interoceptive control signal mediating brain-to-cardiovascular communication. Performance on simulated ground-truth data provides a primary validation of the proposed scheme. Beyond single-subject performance, the estimator generalizes across unseen subjects and datasets, confirming that the neural component does not act as a standalone function approximator. In practice, this supports the use of the most stable source models --- those associated with low mean and low dispersion in Figure \ref{fig:nrmse_violins} --- as reusable initializations for unseen subjects, while the structured failures highlighted in Figure \ref{fig:worst_offdiag} point to the importance of augmenting the model to account for a missed degree of freedom: stroke volume. These conclusions are reinforced by the robustness and comparison analyses. The ablation study showed that removing the left-invertibility terms from the training objective can induce clear estimation failures even in self-prediction, confirming that these constraints are not merely auxiliary regularizers but essential ingredients of the inverse-learning design. Conversely, the comparison with a sliding-mode observer confirmed that satisfactory output fidelity on $H$ may coexist with non-physiological recovery of $P_{sp}$, highlighting the limitations of purely model-based inference under restrictive dynamical assumptions. Taken together, these results support the proposed framework as a relevant compromise between purely data-driven flexibility and purely model-based inference under restrictive assumptions. Future work will focus on improving robustness to noise and on extending the formulation to delayed systems, which is an essential next step both for physiological modeling and for left-invertibility analyses in such settings.
\section*{Availability of data}
\label{sec:AVAILABILITY}
\begin{sloppypar}
\noindent Our dataset is available at: \url{https://musaelab.ca/pass-raw-dataset/}.
\end{sloppypar}

% \section*{Acknowledgements}
% Giuseppe Alessio D’Inverno’s postdoctoral research was funded by the EXCELLENCES/SPRINGBOARD project (RD 2025-04 – Claire Boyer). Giuseppe Alessio D'Inverno would like to thank INdAM-GNCS.
\bibliographystyle{abbrv}
\bibliography{references}  %%% Uncomment this line and comment out the ``thebibliography'' section below to use the external .bib file (using bibtex) .





% \end{thebibliography}

\end{document}